\documentclass[article,12pt]{elsarticle}

\usepackage{graphicx}
\usepackage{amssymb}
\usepackage{lineno}
\usepackage{amsmath}
\usepackage{nicefrac}
\usepackage[utf8]{inputenc}
\usepackage[english]{babel}
\usepackage{natbib}

\newcommand{\m}[1]{\mathbf{#1}} 

\journal{Micron}

\makeatletter
\def\ps@pprintTitle{%
  \let\@oddhead\@empty
  \let\@evenhead\@empty
  \let\@oddfoot\@empty
  \let\@evenfoot\@oddfoot
}
\makeatother

\usepackage{fancyhdr}
\begin{document}

\begin{frontmatter}

\title{Extraction of physically meaningful endmembers from STEM  spectrum-images combining geometrical and statistical approaches}

\author{Pavel Potapov$^{1,*}$ and Axel Lubk$^{1,**}$}

\address{
$^1$ Leibniz Institute for Solid State and Materials Research (IFW), Dresden, Germany

$^*$ p.potapov@ifw-dresden.de

$^{**}$ a.lubk@ifw-dresden.de
    }

\begin{abstract}
This article addresses  extraction of physically meaningful information from STEM EELS and EDX spectrum-images using methods of Multivariate Statistical Analysis. The problem is interpreted in terms of data distribution in a multi-dimensional factor space, which allows for a straightforward and intuitively clear comparison of various approaches. A new computationally efficient and robust method for finding  physically meaningful endmembers in  spectrum-image datasets is presented. The method combines the geometrical approach of Vertex Component Analysis with the statistical approach of Bayesian inference. The algorithm is described in detail at an example of EELS spectrum-imaging of a multi-compound CMOS transistor.   
\end{abstract}


\end{frontmatter}


\section{Introduction} \label{sec:1}
\label{S:Int}

Scanning Transmission Electron Microscopy (STEM) in combination with Electron Energy-Loss Spectroscopy (EELS) or X-rays Energy-Dispersive  Spectroscopy (EDX) allows for \textit{spectrum-imaging}, i. e., high-spatial-resolution STEM imaging, where each pixel is equipped with a spectrum.  Modern STEM instruments now routinely deliver large spectrum-images consisting of tens of thousands of pixels and thousands of energy channels. This opens the possibility to apply the methods of Multivariate Statistical Analysis (MSA) for improving the quality and usability of results. MSA is looking  for statistical correlations in datasets in order to represent data in the more compact, less noisy  and better interpretable form. 

The MSA family includes a number of techniques based on a multitude of approaches and underlying algorithms, like Principal Component Analysis (PCA) \cite{Pearson1901, Hotelling1933, Tipping1997, Jolliffe2002, Malinowski2002,Jolliffe2016}, Independent Component Analysis (ICA) \cite{Comon1994, Hyvarinen2001, Hyvarinen2013}, spectral unmixing \cite{Tarantola1982, Plaza2004, Martinez2006, Bioucas-Dias2012} as well as closely related Multivariate Curve Resolution (MCR) \cite{Tauler1995, Ruckebusch2013,Lavoie2016}, and many others. A common principle of the MSA treatment can be understood when viewing the data distribution in a special space sometimes referred to as a spectral or a factor space \cite{Malinowski2002}. This space typically exhibits a very high dimensionality. For instance, if input spectra are composed of 1000 energy or wave-length channels then the resulting factor space would be a 1000-dimensional space. The key idea of all MSA approaches is the reduction of the factor space dimensionality down to a reasonably small size, which allows to denoise and compress data, or to extract some important variation trends. Still, the number of space dimensions might be noticeably high. Thinking in such a multi-dimensional space is not easy but fortunately some important concepts can be readily picked up from the consideration of particular 2D projections of the data distribution.   

\begin{figure}[!ht]
\centering\includegraphics[width=1.0\linewidth]{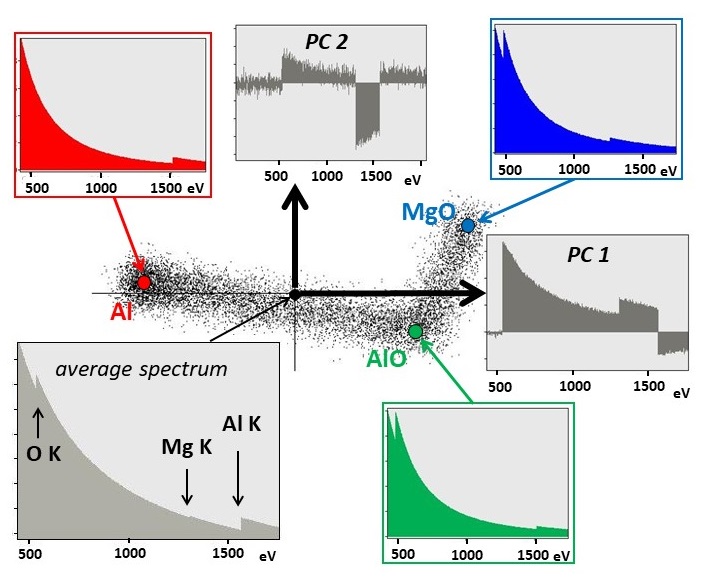}
\caption{Data variations in the synthetic Al-AlO-MgO object shown in the projection onto the plane formed by the major components. The grey-filled insets represent the mean spectrum and its spectral variations along the vectors of the PCA basis  (\textit{PC 1} and \textit{PC 2}). The red, green and blue points indicate the positions of the pure Al, AlO and MgO compounds in  factor space while the insets of  corresponding colors are their spectral signatures.}
\label{fig:1}
\end{figure}

For illustrative purposes, we will consider the  EELS spectrum-image of the synthetic object introduced in \cite{Potapov2016}. Fig. \ref{fig:1} represents the continuous data variations among three compounds - pure Al, AlO and MgO. The  variations  are shown in the projection from the high-dimensional factor space onto the plane formed by the two major principal components found by PCA. This figure is quite instructive to understand the difference in data representation between the various MSA techniques. The red, green and blue points represent the positions of the Al, AlO and MgO compounds in  factor space, while the insets of the corresponding colors are their spectral signatures. The purpose of  spectral unmixing or MCR methods is to find such points, called \textit{endmembers}. All data points are then represented as a linear combination of these endmembers. 

In contrast, PCA, ICA and Varimax methods operate with the \textit{directions} in  factor space and therefore can be referred to as \textit{rotation} methods. They attempt to find  special directions, which represent most significant data variations. The criteria for  significance might be quite different.  For example, PCA searches for the maximal variance in data while ICA  maximizes the non-Gaussian behaviour in the data distribution.    Fig.  \ref{fig:1} shows the spectral signatures (grey) of two major PCA directions in the considered example. They are not spectra of clear physical meaning but differential spectra  with respect to the mean spectrum that is displayed in the lower-left inset\footnote{In this paper we use a \textit{centered} variant of the PCA treatment where the mean spectrum is subtracted from the data prior the PCA decomposition. Similar results can be obtained with non-centered PCA although the dimensionality of the resulting factor space will be increased by one.}. Such differential spectra might be positive or negative depending on whether a given feature is strengthened or weakened when following a certain direction in  factor space. From Fig. \ref{fig:1}, one can immediately notice that the principal components are quite abstract - the data variation can be compactly described using the PCA  basis  but the vectors composing the basis are typically not interpretable. The directions of maximal data variations generally do not coincide with physically meaningful variation trends \cite{Potapov2016}. The other rotation methods might eventually generate more meaningful directions in  factor space. For instance, the combination of PCA and Varimax rotation might under certain conditions result in physically meaningful output \cite{Kotula2003, Keenan2009, Lucas2013}. Still, interpretability of any rotation methods is difficult as they typically operate with an orthogonal basis while the real variation trends commonly do not satisfy the orthogonality restriction.      

On the contrary, the  spectral unmixing problem is formulated as finding some special points in  factor space that probably represent real  compounds existing in a given dataset. The geometrical approaches - N-FINDER \cite{Winter1999}, Vertex Component Analysis (VCA) \cite{Nascimento2005a}, Minimum Volume Simplex Analysis (MVSA) \cite{Craig1994} and many others - consider the problem in terms of  identifying a multi-dimensional simplex that includes all available data points in  factor space. Statistical methods \cite{Moussaoui2006, Dobigeon2009, Arngren2011} explore spectral unmixing as a statistical inference problem. Most of these techniques are used in the area of remote sensing although there are examples of their successful application  in STEM  spectrum-imaging \cite{Dobigeon2012,Shiga2016,Braidy2019}. These decomposition algorithms often put the constraint of positive values on the output endmembers and therefore can be referred to as non-negative matrix factorization methods \cite{Paatero1994}. 

The above consideration suggests that the unmixing approach is preferable as it yields interpretable results with a clear physical meaning. Unfortunately, the unmixing algorithms are usually  more complicated than those utilized  in rotation methods like PCA.  Therefore, many endmembers techniques (see for instance \cite{Kotula2003, Keenan2009, Dobigeon2009, Shiga2016}) use a combination of rotation and unmixing methods - the dimensionality of a dataset is first reduced by rotation and then  relevant endmembers are identified in the reduced space. We will follow the same route and first reduce  the dimensionality of datasets with PCA. The dimensionality reduction with ICA seems less efficient because of higher  computation costs and  a violation of some assumptions on data independence \cite{Nascimento2005b, Bonnet2005}, which  makes ICA less successful in materials science in comparison to, e.g., the area of speech recognition.  

The purpose of the present work is to develop a simple, robust and computation efficient method for extracting endmembers from typical STEM EELS and EDX spectrum-images. Although a number of methods were suggested previously (see for example \cite{Yamazaki2011, Dobigeon2012, Shiga2016}), they often suffer from over-complexity and slow algorithm convergence.  In our method we will combine the geometric approach with statistical Bayesian inference. The latter will be used in its simplified form not requiring any costly optimisation like application of stochastic Markov chains as in \cite{Dobigeon2009}. The extraction of endmembers will be performed in the factor space preliminary reduced with PCA. We assume that the optimal PCA dimensionality is already determined with, for instance screeplot \cite{Cattel1964} or anisotropy  \cite{Potapov2019} methods. Furthermore, this paper will be limited to the \textit{linear} decomposition problem. The influence of non-linearity on the spectra formation will be  considered in Discussion (Section \ref{sec:4}).

The paper is organized as follows. Section \ref{sec:2} describes an exemplary spectrum-image that is used to explain the workflow in detail. Section \ref{sec:2.1} provides some pre-acquired information about the composition of the object needed for cross-validation of the forthcoming  results. Section \ref{sec:3.1} describes the details of the dimensionality reduction with PCA.   Sections \ref{sec:3.2} and  \ref{sec:3.3} introduce the classical VCA method and its modified multi-run variant. We then suggest to combine this kind of the VCA treatment with the Bayesian inference (Section \ref{sec:3.4.1}). The filtration and clustering of the obtained results plays an important role in our algorithm  as explained in Section \ref{sec:3.4.2}. Section \ref{sec:3.4.3} describes the calculation of endmember spectra and abundances. The results of the treatment are shown and analysed in Section \ref{sec:3.5}. The restrictions and applicability of the suggested method are discussed in Section \ref{sec:4}. 

\section{Exemplary EELS spectrum-image} 
\label{sec:2}

\subsection{Object for spectrum-imaging} 
\label{sec:2.1}

\begin{figure}[ht]
\centering\includegraphics[width=0.7\linewidth]{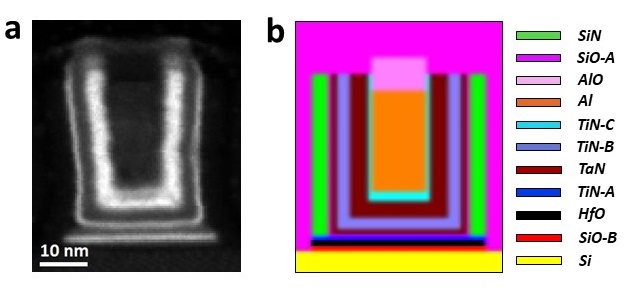}
\caption{The CMOS transistor examined in the present paper: (a) shows its HAADF image and (b) depicts schematically the compounds expected to be observed in the transistor. The composition of the compounds is listed in Table \ref{tab:1}.}
\label{fig:2}
\end{figure}

A modern CMOS transistor was chosen as a model object to test the designed MSA procedure. The transistor consisted of a number of nano-scale layers manufactured in order to optimize the speed, switching potential and leakage current of the device \cite{Potapov2013}. Fig. \ref{fig:2} shows a High-Angle Annular Dark Field (HAADF) image of the transistor as well as a schematic depiction of its  constituting compounds. This picture was deduced from the combination of the different analytical methods - STEM EELS/EDX, Auger spectroscopy and ToF-SIMS. In total, the 11 various  compounds were identified, although some   compounds differed in composition only marginally as seen from Table \ref{tab:1}. 

\begin{table}[h]
\centering
\begin{tabular}{l l l}
\hline
\textbf{Compound notation} & \textbf{Composition (at.\%)} \\
\hline
Si & 100\% Si \\
SiO-A & 33\% Si - 67\% O \\
SiO-B & 29\% Si - 57\% O - 14\% N\\
HfO & 33\% Hf - 67\% O \\
TiN-A & 50\% Ti - 50\% N \\
TiN-B & 50\% Ti - 40\% N - 10\% O \\
TiN-C & 45\% Ti - 45\% N - 10\% Al \\
TaN & 50\% Ta - 50\% N \\
Al & 80\% Al - 20\% Ti \\
AlO & 30\% Al - 10\% Ti - 60\% O \\
SiN & 43\% Si - 57\% N \\

\hline
\end{tabular}
\caption{Composition of compounds constituting the investigated CMOS transistor.}
\label{tab:1}
\end{table}

\subsection{Experimental conditions}
\label{sec:2.2}

The TEM cross section of the CMOS device was prepared by FIB at 30 kV followed by Ga ion milling at 5 kV. The final thickness of the sample was approximately 50 nm. The region of interest was then scanned in the Titan G2 S/TEM microscope operated at 300kV and equipped with a Gatan Quantum energy filter. The probe size was about 0.2 nm with a beam current of 120 pA and a dwell time of 30 ms. The scan comprised of $98 \times 118$ pixels covering an area of $43 \times 52$ nm and 2048 channels spanning an energy-loss range of 50...2098 eV.

\section{Endmembers Algorithm} \label{sec:3}
\subsection{Dimensionality reduction with PCA} \label{sec:3.1}

As mentioned in Section \ref{sec:1}, the dimensionality of the data should be first reduced by applying the PCA procedure. A dataset is represented by an $m \times n$ matrix $\m{D}$,  where spectra are placed on the matrix rows and each row corresponds to an individual STEM probe position\footnote{Although STEM probes may be originally arranged  in  1D (linescan), 2D (datacube) or in a configuration of higher dimensionality, they can be always recast into a 1D column as the spatial correlation among probes does not play any role in the PCA treatment.}. Accordingly $m$ is the number of STEM probes and $n$ is the number of energy channels. PCA decomposes the data as
\begin{equation}
\m{D} = \m{T  P\textsuperscript{T}}
\label{eq:1}
\end{equation} 

\noindent
where $\m{P}$ is an $n \times n$ \textit{loading} matrix describing principal component spectra and $\m{T}$ is an $m \times n$ \textit{score} matrix showing the contribution of components in the dataset.

PCA is  an efficient way to reduce the dimensionality of data corrupted by Gaussian noise. The dominant noise in STEM spectrum-imaging is, however, Poissonian; therefore the classical PCA is not directly applicable. Supplementary Material 1 demonstrates a typical issue arising from Poisson noise in EELS spectrum-images. The problem is addressed by rescaling data prior PCA such that the noise variance is equalized across a dataset.  One  common rescaling strategy is the Anscombe transformation that converts  a random variable with a Poisson distribution into one with an approximately  Gaussian distribution \cite{Anscombe}. Another approach that was widely-used in the last decades for STEM EELS and EDX data is the optimal weighting after Keenan and Kotula \cite{Keenan2004}  prior to PCA. For the considered dataset, they lead to very similar results as demonstrated in Supplementary Material 2. For the sake of consistence with previous publications, this article rescales the data as in \cite{Keenan2004}:

\begin{equation}
\m{\widetilde{D}} =  \m G \textsuperscript{-1/2} \m{D}  \m H \textsuperscript{-1/2}
\label{eq:2}
\end{equation}

\noindent
where $\m H \textsuperscript{-1/2}$ is an $n \times n$ diagonal matrix with the inverse square root mean spectrum (i.e., averaged over all probe positions) on the diagonal and $\m G \textsuperscript{-1/2}$ is an $m \times m$ diagonal matrix with the inverse square root mean image (i.e., averaged over all energy channels) on the diagonal.

Accordingly, the weighted dataset is decomposed by PCA into

\begin{equation}
\m{\widetilde{D}} = \m{\tilde{T}  \tilde{P}\textsuperscript{T} } 
\label{eq:3}
\end{equation} 

\noindent
where the score and loading matrices $\m{T}$ and $\m{P}$ are replaced for their weighted analogs $\m{\tilde{T}}$ and  $\m{\tilde{P}}$.  Optionally, data can be also centered, i.e., the mean spectrum can be subtracted  from  the dataset after the weighting pre-treatment. 

The columns of loading matrix $\m{\tilde{P}}$ represent weighted spectra of principal components expressed in the original energy channels. It is important to sort the principal components in the order of their significance. In PCA, the components are ranked according their variance, i.e., the variance of the data along the corresponding column of $\m{\tilde{T}}$. At the next step,  the dimensionality of the dataset is reduced by truncating the number of principal components as:

\begin{equation}
\m{\widetilde{D}} \approx {\m{\tilde{T}}_k  \m{\tilde{P}}_k\textsuperscript{T}}
\label{eq:4}
\end{equation} 

\noindent
where index $k$ indicates that only the $k$ first columns in $\m{\tilde{T}}$ and $\m{\tilde{P}}$ are retained while the rest is removed. In other words, energy dimensionality  is reduced from $n$ to $k$. The choice of $k$ is a complicated task that was addressed numerous times based on different argumentation. In the present work, we use the screeplot approach \cite{Cattel1964} cross-validated with the recently invented  anisotropy method \cite{Potapov2019}.

\begin{figure}[ht]
\centering\includegraphics[width=1.0\linewidth]{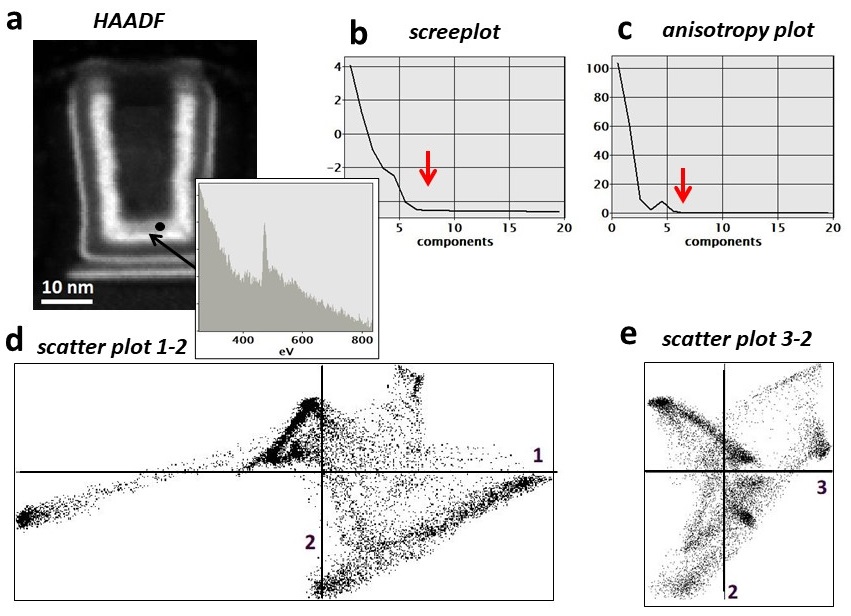}
\caption{PCA results for the object described in Section \ref{sec:2}. (a) shows the HAADF image acquired during spectrum-imaging and the inset is a typical EEL spectrum obtained in the course of acquisition. (b) and (c) show  screeplot and anisotropy plot of the extracted PCA components, where the red arrows indicate the cut-off value. (d) and (e) display 2D projections of the data distribution on planes formed by (d) the 1st and 2nd, and  (e) the 2nd and 3rd principal components, respectively.} 
\label{fig:3}
\end{figure}

Fig. \ref{fig:3} shows the PCA results for the spectrum-image described in  Section \ref{sec:2}. Both the screeplot and anisotropy methods suggest to retain 7 major principal components ($k=7$) as  representing all relevant variations in the data. This is a dramatic reduction compared to the original dimensionality of 2048 energy channels. 

 The projections of the data distribution from the 7-dimensional factor space onto selective planes formed by the principal components are very instructive to explain the further MSA treatment. These projections, also referred to as \textit{scatter plots}, are displayed in Fig. \ref{fig:3} d,e reflecting the complexity of the data variation in the considered dataset. We stress that this gives only a general idea about the geometry of the data variation as its full representation would require a visualisation in 7-dimensional space, which is rather difficult to achieve.

\subsection{Classical VCA treatment} \label{sec:3.2}

 The next task is to identify special limit points in factor space, so called endmembers, which can completely describe the variations in factor space and represent certain chemical compounds constituting a given dataset.  
 
 We will formulate this task not in the original $n$-dimensional spectral space but in the PCA-reduced $k$-dimensional space. In this space, the data points are cast in the coordinates of $k$ major principal components and defined by the  matrix $\m{\tilde{T}}_k$. 
 
In the absence of noise, the linear mixing model implies that

\begin{equation}
\m{\tilde{T}}_k\textsuperscript{T} = \m{M} \m{A}
\label{eq:5}
\end{equation} 

\noindent
where    $\m{M}$ is a $k \times r$   matrix consisting of the position of $r$ endmember points expressed in $k$ major principal components and $\m{A}$ is an $r \times m$  matrix containing the contributions of the endmembers in the data points.  Note that the data distribution in the left part of (\ref{eq:5}) are represented by transposed $\m{\tilde{T}}_k$ in order to be consistent with the common conventions used in spectra unmixing, e.g. in \cite{Nascimento2005a}.

The purpose of the geometrical unmixing methods is to incorporate  all available data points inside  a certain simplex defined by the set of the endmembers (columns of  $\m{M}$). In particular, VCA attempts to find such endmembers by identifying the data points with the most extreme positions in the $k$-dimensional factor space.

\begin{figure}[ht]
\centering\includegraphics[width=0.6\linewidth]{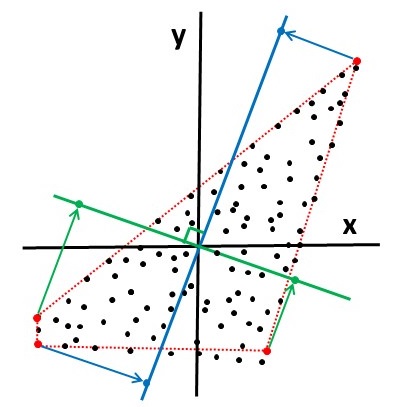}
\caption{Schematic explanation of the classic VCA algorithm for a two-dimensional case. The data points are projected on the random (blue) line and the extremities of the projection are found. Another line (green) perpendicular to the first one is drawn and the procedure is repeated. The resulting extreme points (red) form a polygon, which is supposed to include all available data points.}
\label{fig:4}
\end{figure}

The basic idea of VCA can be understood from  Fig. \ref{fig:4}, where a simplified two-dimensional situation is sketched. The factor space might be two-dimensional from the beginning or be reduced to the 2D space by PCA; in the latter case the $x$ and $y$ axes represent the 1st and 2nd principal components. VCA was originally designed for hyperspectral remote surveillance \cite{Nascimento2005a}, where  conical projections were utilized to account for the illumination variability due to the surface topography. In the present paper, we describe VCA in the slightly modified form adapted for application in STEM spectrum-imaging.  

VCA starts with generating a line of random orientation in  factor space. All data points are projected onto this line and the point with the most extreme projected position is chosen. In the original formulation of VCA \cite{Nascimento2005a}, only one extreme point is selected. We note, however, that  both extremal points located at the opposite ends of the projection line can be  utilized as shown in Fig. \ref{fig:4}. This strategy is reasonable for cases when the maximal number of extreme points has to be generated within the shortest time. 

In the next step, the second line perpendicular to the first one is plotted and the next extremal data points are identified. The procedure is repeated until all dimensions of the probed space are exhausted (2 dimensions for the example in Fig. \ref{fig:4}).  This results in the set of endmembers, which typically form a simplex that includes all available data points.

\subsection{Scattered VCA treatment} \label{sec:3.3}

As seen from Fig. \ref{fig:4}, the incorporation of all data points in the VCA simplex is not guarantied. Depending on the choice of the first (randomly oriented) line,  a significant part of the data might remain unattended. Another issue of the classical VCA is the uncertainty in the number of endmembers. The described procedure  generates $k$ or 2$k$ endmembers depending on the treatment of extremal points  while the natural choice for the $k$-dimensional space would be a simplex with $(k+1)$ corners.
    
To address these issues, a repeatable application of VCA was suggested \cite{Spiegelberg2017}. The procedure sketched in Fig. \ref{fig:4} is repeated a number of times and all the obtained endmembers are then clustered to extract some reasonable \textit{mean} endmembers. Clustering can be performed using the $k$-means or any other method allowing for the control of the final number of endmembers. 

Despite of the general fruitfulness of the idea, the approach of \cite{Spiegelberg2017} shows certain limitations: i) a significant ratio of the generated endmembers is redundant as demonstrated in Supplementary Material 4, which reduces the statistical significance of their clustering; ii)  it does not account for the noise corrupting the data. The noise might sporadically generate data points situated quite far from the "true", noise-free positions of the endmembers. Although such points are rare, they will be most probably identified as endmembers due to their extreme positions. Thus, the results are expected to be strongly influenced by the noise level in a given dataset.   The algorithm is also not robust against outliers caused by artefacts, e.g.,  X-ray spikes.

\subsection{Scattered VCA treatment combined with Bayesian refinement} \label{sec:3.4}

We therefore introduce another method accounting for the noisy nature of STEM spectrum-images. As explained previously, the weighting pre-treatment effectively converts Poissonian noise into Gaussian noise. The truncated PCA treatment removes most of noise from a dataset, still some fraction of the Gaussian noise remains in the PCA-reduced factor space. In the presence of Gaussian noise, equation (\ref{eq:5}) becomes

\begin{equation}
\m{\tilde{T}}_k\textsuperscript{T} = \m{M} \m{A} + \m \Omega
\label{eq:6}
\end{equation}  

\noindent
where $\m \Omega$ is a $k \times m$ matrix consisting of $m$ digitized realisations of the  $k$-variate Gaussian distribution $\mathcal{N}(\m{M} \m{A},\sigma)$. $\m{M} \m{A}$ represents  hypothetical  noise-free points in the dataset distribution and $\sigma$ is the noise standard deviation, which is expected to be same along all $k$ dimensions.   

The problem of accurately estimating the noise standard deviation $\sigma$ in  PCA-reduced datasets has been addressed several times in  connection with the truncation of the principal components (see for instance \cite{Kritchman2008}). For the purpose of the present paper, $\sigma$ needs not to be determined very precisely, thus even the simplest method based on the estimation of the residual data matrix \cite{Malinowski1977} is sufficient. 

\subsubsection{Bayesian refinement of endmembers in the VCA routine} \label{sec:3.4.1}

\begin{figure}[!ht]
\centering\includegraphics[width=0.7\linewidth]{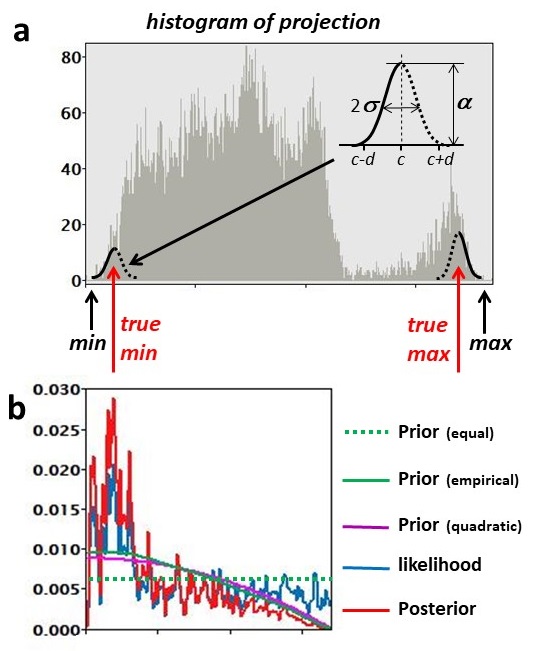}
\caption{(a) Exemplary histogram of the data points projected on the random line and (b) extraction of the endmembers channels by Bayesian inference. The nominal "min" and "max" channels are refined to the true ones using the prior (green) and likelihood (red) distributions. The alternative priors described in Appendix E are also shown.}
\label{fig:5}
\end{figure}

As in the classical VCA, we start with generating a line of random orientation in $k$-dimensional factor space. The projections of all data points onto that line fall within a certain range between the minimal and maximal values. We can build a histogram of the projected points as shown in Fig.\ref{fig:5}a. Suppose that the histogram is composed  of $l$ channels. The naive VCA treatment implies choosing the minimal and maximal channels that still consist of non-zero counts as the output extremities. It is known, however, that the "true", noise-free data distribution along the projection line is smeared out by the Gaussian distribution\footnote{It can be be shown that the multivariate Gaussian distribution converts into the one-dimensional one with the same $\sigma$ parameter after the projection.} due to noise. Therefore the observable counts $X(i)$ at  histogram channels $i=1,...,l$ should approach the convolution of true counts $Y(i)$ with a sampled Gaussian kernel $S(j)$:   

\begin{equation}
X(i) \approx \sum_{j=-\infty}^{j=+\infty}Y(j) S(j-i)
\label{eq:7}
\end{equation}

\noindent
where $S(j) = 1/\sqrt{2\pi \sigma^*} \exp(- j^2/ 2\sigma^{*2})$ with $\sigma^*$ being the noise standard  variation expressed in the number of channels needed to capture the $\sigma$ interval. As the Gaussian kernel $S(j)$ decays quickly with $ \lvert j \rvert$,  the summation limits in (\ref{eq:7}) can be set to $\pm d$ where $d$ is around $3\sigma^*$.  

The task is now to determine at which histogram channels the "true" (not smeared out by noise) extremities  are most probably located. We will evaluate  that using Bayesian inference. 

Consider for brevity only the left part of the histogram in Fig. \ref{fig:5}a. We introduce a random variable $C$ pointing to the position of the  "true" extremity and evaluate the probability that the extremity is situated at a given channel $c$, or in other words, that $C=c$. As a Bayesian prior, the equal probability distribution can be used: 
\begin{equation}
\mathcal{P}[C=c] = \frac{1}{l}
\label{eq:8}
\end{equation}

\noindent
where $l$ is, as before, the number of channels in the histogram.

The next step of the Bayesian inference is the calculation of the likelihood for a given hypothesis. Formula (\ref{eq:7}) suggests that if channel $c$ is indeed the left edge of the true noise-free data scattering, then all channels to the left from $c$ should be populated by the counts closely resembling the tail of Gaussian kernel $S(j)$. Note that we consider only the \textit{left} tail of Gaussian kernel $\mathcal{S}(j)$ centered at $c$. The right tail is strongly affected by the unknown noise-free distribution $Y(j)$ and cannot be used in the analysis. Our approximation implies also that $Y(j)$ at the right side from $c$ does not change significantly the histogram shape at the left side from $c$. Appendix C shows that this assumption is approximately fulfilled provided that $Y(j)$ varies smoothly.  

The possible deviations of the observable counts $X(i)$ from the kernel $S(j)$ shape due to random noise variation can be evaluated by the standard likelihood estimation. As above, we truncate $S(j)$ in the limits $\pm d$, thus only the range $[c-d,c]$ is considered.  Then the likelihood of the hypothesis that $C=c$ is

\begin{equation}
\mathcal{P}(X_{c-d},...X_c|C=c) = \prod_{i=c-d}^{c}\frac{1}{\sqrt{2\pi V}}  \exp{\left (- \frac{(X(i)-\alpha(c) S(c-i))^2}{2V}\right)}
\label{eq:9}
\end{equation}

\noindent
where $V$ is the variance of $C$ and $\alpha(c)$ is the unknown value of "true", noise-free count at channel $c$.  In the present work, we estimate $\alpha(c)$ by equating the sum of $X(i)$ and $\alpha(c) S(c-i)$  in the $[c-d,c]$ range. This intuitively clear assumption maximizes  the likelihood $\mathcal{P}$ as proven in Appendix D. 

Random variations due to the finite number of counts must follow the Poisson statistics with the variance equalling the noise-free value ($V = \alpha(c) S(c-i)$). Therefore, formula (\ref{eq:9}) can be rewritten as 

\begin{equation}
\mathcal{P}(X_{c-d},...X_c|C=c) = \prod_{i=c-d}^{c}\frac{1}{\sqrt{2\pi}} \frac{1}{\sqrt{\alpha(c) S(c-i)}} \exp{\left (- \frac{(X(i)-\alpha(c) S(c-i))^2}{2\alpha(c) S(c-i)}\right)}
\label{eq:10}
\end{equation}

According to Bayes theorem, the  posterior probability that $C=c$ is finally calculated as the prior times the likelihood divided by the sum of the likelihoods for all considered hypothesis:

\begin{equation}
\mathcal{P}(C=c|X_{c-d},...X_c) =  \frac{\mathcal{P}(X_{c-d},...X_c|C=c)}{\sum_{i=1}^{l}\mathcal{P}(X_{c-d},...X_c|C=i)} \mathcal{P}[C=c]
\label{eq:11}
\end{equation}
\noindent

The resulting probability distribution allows to evaluate not only the position of the "true" extremity but also its confidence range. The present version of our code, however, does not use this information but simply determines the most probable $c$ providing the maximum of $\mathcal{P}$. The statistical significance of each obtained endmember position will be later implicitly taken into account during the clustering stage in Section \ref{sec:3.4.2}   

The described procedure determines a certain histogram channel, not the position of the potential endmember. The required  endmember is then found by averaging the positions of all data points falling into the target channel. As seen from Fig. \ref{fig:5}a, this typically involves averaging of at least 10 data points.

We found that the procedure delivers a quite realistic set of potential endmembers although, in few cases, the generated endmembers appear to be located deeply inside the region of the data distribution, not at its ends.  This can happen if the shape of the histogram somewhere in its middle part eventually mimics the shape of a Gaussian kernel. To suppress such outliers, we explored alternative to (\ref{eq:8}) priors. These alternative priors are described in Appendix E and shown in Fig. \ref{fig:5}b. 

\subsubsection{Clustering of scattered endmembers} \label{sec:3.4.2}

\begin{figure}[p]
\centering\includegraphics[width=1.0\linewidth]{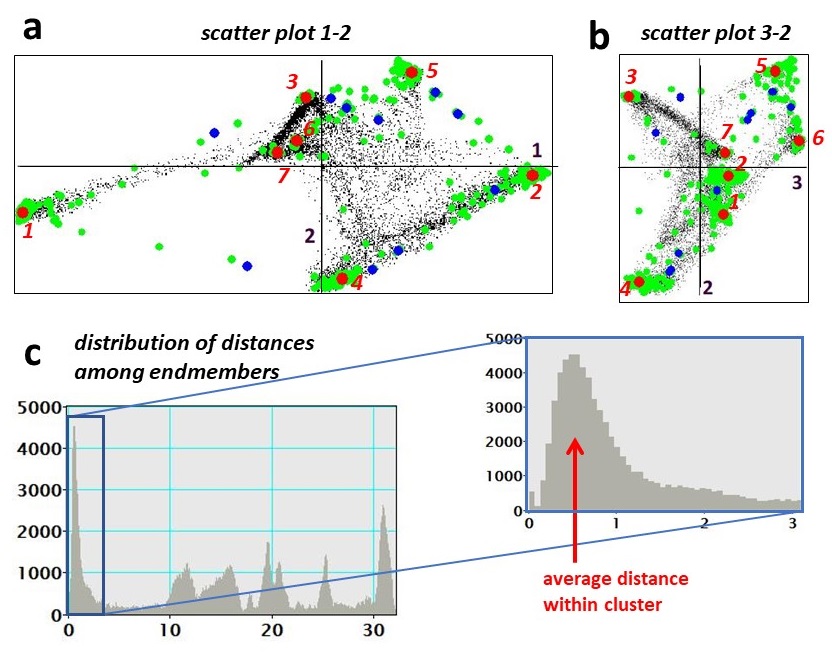}
\caption{Clustering of the obtained potential endmembers (green spots) by the mean-shift method. The endmembers are shown in 2D projections on planes formed by (a) the 1st and 2nd, and  (b) 2nd and 3rd principal components. (c) shows the histogram of the mutual distances among  the obtained endmembers, which allows to determine the average distance among the  endmembers within a cluster. The blue and red spots show the output of the mean-shift method, i.e., the  positions of the maximal density of the potential endmember distribution. According to their ratings (Table \ref{tab:2}), part of these positions (red) are retained as a final endmember set while the remainders (blue) are neglected. The retained positions are indexed as listed in Table \ref{tab:2}.}
\label{fig:6}
\end{figure}

As in the classical VCA, there is still an uncertainty related to the random choice of the initial line in factor space. We solve this problem by repeating several times the VCA procedure as suggested in \cite{Spiegelberg2017}.  In the considered example we ran the VCA procedure 40 times, which yielded $40 \times 7(\mathrm{dimensions}) \times 2(\mathrm{extrema}) = 560$ potential endmembers.  Fig. \ref{fig:6}a,b shows the positions of these endmembers (green points) as seen in the different 2D projections (keep in mind that the total number of dimensions is 7). Notably, most of the endmembers tend to be clustered around certain positions. The next step of the processing is the determination of these positions by an appropriated clustering technique \cite{Ghuman2017}.  

One option is the k-means clustering method employed in \cite{Spiegelberg2017}. This method, however, requires the number of endmembers to be pre-defined for a given dataset and is also sensitive to the heavy  outliers.  As an alternative, we have employed a mean-shift clustering algorithm \cite{Fukunaga1974}, which is searching for the most dense areas in the distribution of the obtained potential endmembers. 

The algorithm starts with evaluation of the average distance among the available potential endmembers (Fig. \ref{fig:6}c). There are several peaks in the histogram of mutual distances indicating that several distinct clustering centers exist in the considered space. The most left peak (zoomed in Fig. \ref{fig:6}c) reflects the distribution of the potential endmembers around a \textit{nearest} clustering center and the positions of its maximum characterizes the average distance among the  endmembers within the clusters.  

We outline a hyper sphere around each potential endmember with the radius corresponding the position of the first maximum in the histogram of the mutual distances (red arrow in Fig. \ref{fig:6}c).   Overlapping spheres are eliminated by comparing the number of neighbouring endmembers included in each sphere, what will be called \textit{rating}. The endmembers with lower ratings are removed while the "winning" endmembers gain the ratings of all eliminated competitors.

\begin{table}[h]
\centering
\begin{tabular}{l l l}
\hline
\textbf{endmember} & \textbf{rating} & \textbf{associated compound} \\
\hline
1 & 28752 & \textbf{Si} \\
2 & 6952 & \textbf{TiN} \\
3 & 859 & \textbf{SiO} \\
4 & 810 & \textbf{TaN /HfO} \\
5 & 766 & \textbf{AlTiO} \\
6 & 75 & \textbf{AlTi} \\
7 & 69 & \textbf{SiN} \\
8 & 22 & \textbf{?} \\
9 & 6 & \textbf{?} \\
10 & 4 & \textbf{?} \\
11 & 2 & \textbf{?} \\
12 & 2 & \textbf{?} \\
13 & 2 & \textbf{?} \\
14 & 2 & \textbf{?} \\
15 & 2 & \textbf{?} \\
16 & 2 & \textbf{?} \\
17 & 2 & \textbf{?} \\

\hline
\end{tabular}
\caption{The ratings of the endmembers obtained by the procedure described in section  \ref{sec:3.4.2}. The last column shows the possible identification of the endmembers with the real compounds present in the device.}
\label{tab:2}
\end{table}

Then, the mean position of the endmembers within each remaining sphere is calculated and the sphere center is shifted to that position. In this way, only spheres in  areas with a high endmembers density are retained and their centers shift towards the place of maximal density. The algorithm typically converges after two or three iterations. 

In the considered example, the described procedure eliminates most of the original potential 560  endmembers and retains only 17 indicated by the red and blue points in Fig. \ref{fig:6}a,b. Their accumulated ratings vary dramatically as seen from Table \ref{tab:2}.  It is evident that half of these endmembers represent just few outlier events. Seven endmembers, depicted as numbered red points in Fig. \ref{fig:6}a,b, show a clear statistical significance.   The other endmembers (blue in Fig. \ref{fig:6}a,b) are neglected. Endmember 8 is a questionable case as it might be considered as relevant according the rating criterion. Inclusion or exclusion of  member 8 from the set of relevant endmembers will be discussed in Section \ref{sec:4.2}.  

\subsubsection{Calculation of endmember spectra and abundances} \label{sec:3.4.3}

In Section \ref{sec:3.4.2}, the endmembers were determined according to their positions in the PCA reduced $k$-dimensional space, which defines matrix $\m M$ introduced in (\ref{eq:5}). Equation (\ref{eq:4}) can be then represented as:

\begin{equation}
\m{\widetilde{D}} \approx {\m{A}\textsuperscript{T}\m{M}\textsuperscript{T}  \m{\tilde{P}}_k\textsuperscript{T}}
\label{eq:12}
\end{equation} 

Formula (\ref{eq:12}) ignores the  noise term $\Omega$ introduced in (\ref{eq:6}). Thus, our approach accounts for $\Omega$ in finding $\m M$ but neglects noise in evaluating $\m A$. We define now a weighted abundance matrix $ \tilde{\m C} =\m A^T$ and a weighted endmember matrix $ \tilde{\m E} =  \tilde{\m P}_k \m M$. Then, (\ref{eq:12}) becomes

\begin{equation}
\m{\widetilde{D}} \approx {\m{\tilde{C}}  \m{\tilde{E}}\textsuperscript{T}}
\label{eq:13}
\end{equation} 

To calculate $\tilde{\m C}$, we rewrite (\ref{eq:13}) as $m$ systems of $n$ linear equations.

\begin{equation}
\m{\tilde E} \m{\tilde C}\textsuperscript{T} = \tilde{\m D} {\textsuperscript{T}}
\label{eq:14}
\end{equation}

\noindent
and search for the contribution of $r$ endmembers in each of $m$ pixels. As $r\ll n$, these systems are over-determined and can only be solved in the least square sense. However the solution might be unstable. To stabilise the solution we used Tichonov regularization \cite{Tichonov}:

\begin{equation}
\tilde{\m C}\textsuperscript{T} = ( \tilde{\m E} {\textsuperscript{T}}\tilde{\m E} +\lambda \m I) \textsuperscript{-1} \m {\tilde E\textsuperscript{T}} \m {\tilde D\textsuperscript{T}}  
\label{eq:15}
\end{equation}

\noindent where $\m I$ is a $r \times r$ unity matrix and $\lambda$ is a regularization parameter, which should be much less than the typical values in $\tilde{\m D}$. In the present work, it was taken equal to $10^{-5}$ of the $\tilde{\m D}$ mean value.  

Formula (\ref{eq:14}) represents unmixing in the weighted factor space while spectra in the original spectral space are finally needed. This requires rescaling  the data back to the original scale, opposite to what was performed  in (\ref{eq:2}):

\begin{equation}
\m{D} =  \m G \textsuperscript{1/2} \m{\widetilde{D}}  \m H \textsuperscript{1/2} 
\label{eq:16}
\end{equation}

\noindent
where $\m H \textsuperscript{1/2}$ is an $n \times n$ diagonal matrix with the  square root mean spectrum (spectra averaged over all probe positions) on the diagonal and $\m G \textsuperscript{1/2}$ is an $m \times m$ diagonal matrix with the  square root mean image (image averaged over all energy channels)   on the diagonal.

Our task is the representation of data matrix $\m{D}$ as

\begin{equation}
\m{D} \approx \m{C} \m{E}\textsuperscript{T}
\label{eq:17}
\end{equation} 

\noindent
where $\m E$ is a $n \times r$ \textit{endmember} matrix representing the $r$ endmember spectra and and $\m C$ is a $m \times r$ \textit{abundance} matrix containing the contributions of the endmembers into the dataset. Formula (\ref{eq:17}) is analogous to (\ref{eq:1}) although endmembers $\m E$ and their abundances $\m C$ are supposed to have a clear physical meaning in contrast to the PCA results. 

Therefore, we define  endmember matrix $\m E$ as

\begin{equation}
\m{E} = g \m H \textsuperscript{1/2} \tilde{\m E} 
\label{eq:18}
\end{equation}

\noindent
where  $g$ is the mean value of the diagonal of matrix $\m G \textsuperscript{1/2}$. With this definition, the endmember spectra are scaled closely to the original average spectra intensity in the sample. From (\ref{eq:13}) and (\ref{eq:16}), abundance matrix $\m C$ is then  defined as 

\begin{equation}
\m{C} =  \frac{1}{g}\m G \textsuperscript{1/2}  \m {\tilde C} 
\label{eq:19}
\end{equation}

 The \textit{sum-to-one} condition  is often required for abundance matrix $\m C$, which means that the sum of abundances at each given pixel should be 1. This can be in principle achieved in weighted abundance matrix $\tilde{\m C}$ by augmenting $\tilde {\m E}$ and $\tilde{\m D}^T$ in (\ref {eq:14}) with an additional row composed of unities.  However, unweighted abundance matrix $\m C$ would then show the variable sum of abundances, e.g. higher in the thicker places of the sample and lower in the thinner ones. If the \textit{sum-to-one} condition is needed, a simpler solution is to replace $\m C$ for $\m C^*$ where each row is normalized such that it sums up exactly to 1.  Note that although $\m C^*$ represents adequately the distribution of endmembers across the sample, it cannot be used for the data reconstruction from (\ref {eq:17}).

\subsection{Results for exemplary EELS spectrum-image} \label{sec:3.5}

\begin{figure}[p]
\centering\includegraphics[width=1.0\linewidth]{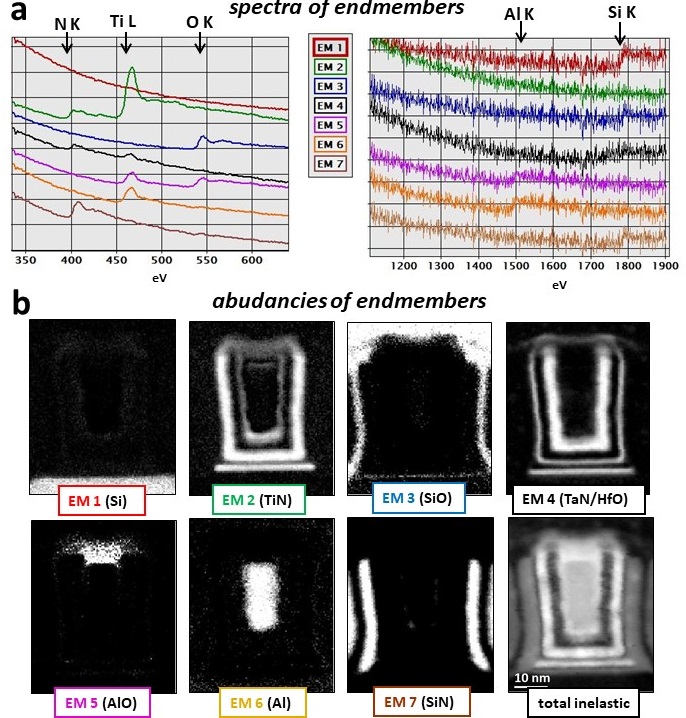}
\caption{Endmembers analysis of the STEM EELS spectrum-image of the CMOS transistor. (a) shows the spectra of the endmembers in the lower (left) and higher (right) energy regions. (b) are the abundances of the extracted endmembers. For comparison purpose, the map of the  EELS inelastic signal integrated over the whole available energy range is shown in the right-lower corner. This map roughly correlates with the density of material.}
\label{fig:7}
\end{figure}

Fig. \ref{fig:7}a shows the spectra of the  7 endmembers extracted in the considered example. The spectra are displayed in the lower and higher energy regions.  The appearance of the characteristic EELS ionisation edges for N, Ti, Al and Si allows to identify the endmembers as the compounds listed in Table \ref{tab:2}. They fit generally the expected compound list in Table \ref{tab:1}, which  demonstrates phenomenologically the validity of the above endmembers algorithm. Note that the characteristic EELS features in the high energy domain are quite weak. That happened because the microscope crossover was not optimised for high-energy EELS \cite{Craven2017}, therefore such features tended to be lost during the standard EELS processing. Nevertheless they appear quite clearly after the described PCA dimensionality reduction followed by the endmembers extraction.

The abundances of the extracted endmembers  are shown in Fig. \ref{fig:7}b in comparison with the map of the total inelastically scattered signal (right-lower corner), which roughly reflects the variation of material density over the considered CMOS device. The abundances of the endmembers fit well the expected distribution of the compounds sketched in Fig. \ref{fig:2}b with one exception. The expected compounds TaN and HfO are not resolved and appear as a single component (endmember 4). This is because Hf and Ta show weak and overlapping EELS edges at high energies (1793eV for Ta and 1716eV for Hf \cite{Egerton}). The spectrum of endmember 4 (black curve in Fig. \ref{fig:7}a, right) indeed demonstrates a smooth increase of the intensity in this region although the individual edges are not separated. This, however, does not mean that they cannot be resolved in principle. The investigation of the same device at higher resolution and more dense sampling allows to discriminate successfully the TaN and HfO compounds \cite{Potapov2013}.     

\section{Discussion} \label{sec:4}

\subsection{Applicability of linear-mixing model} \label{sec:4.0}

The  treatment described in Section \ref{sec:3.4} relies on a linear mixing model. In reality, some deviations from linearity might be introduced either by various physical and instrumental effects or by manipulations occurring in the course of processing. The CMOS device described in Section  \ref{sec:2} offers a good possibility to examine the applicability of the linear-mixing model for EELS spectrum-images. Although it consists of a number of different compounds, the composition variations  are mostly characterized by mixtures of two given compounds while mixtures of three or more compounds are less common. This originates from the CMOS manufacturing process, which involves subsequent deposition of various non-planar layers on top of each other.   Therefore, the variation trends resemble mostly lines, not planes or objects of higher dimension,  connecting certain points in factor space. Inspecting the shape of these lines (i.e., whether they are straight or not) allows to roughly assess the linearity of mixing. The straightness of lines in particular projections is evidently necessary but not a sufficient condition for linearity due to the multi-dimensionality of the data variations. Still, as most variations are captured by the major principal components, such visual inspection is useful.  

The 2D projections in Fig. \ref{fig:6}a,b suggest that the variation trends follow more or less straight lines in most cases. Consider, for instance, endmembers 2 and 4 that represent  spatially close layers of TiN and TaN/HfO respectively (see Fig. \ref{fig:2}). The variation trend between them is clearly seen as a straight line in Fig. \ref{fig:6}a. Another example of a reasonably linear mixture is a straight line between endmembers 3 (SiO) and 7 (SiN). A slight deviation from the linearity can be, however, noticed in the variation trend between endmembers 1 (Si) and 7 (SiN).     

\begin{figure}[ht]
\centering\includegraphics[width=1.0\linewidth]{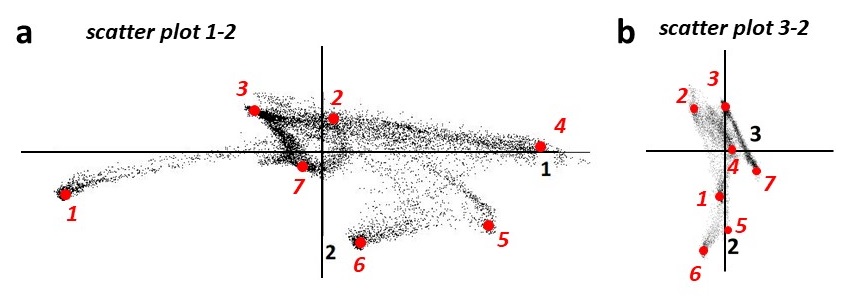}
\caption{Data distribution shown in 2D projections on planes formed by (a) 1st and 2nd, and (b) 2nd and 3rd principal components after PCA \textit{without} the weighting pre-treatment. This figure should be compared with the regular processing flow including the weighting (Fig. \ref{fig:3}d,c).}
\label{fig:8}
\end{figure}

It is instructive to evaluate the effect of the weighting pre-treatment on the validity of the linear mixing model.  Fig. \ref{fig:8} shows the data projections on planes formed by the major principal components for the case when PCA was performed \textit{without} the weighting\footnote{In contrast to example in Fig. \ref{fig:1} and Supplementary Material 1, the  PCA is still able to extract roughly few major principal components in this spectrum-image even without weighting. That is because the major meaningful variations are significantly larger than the noise level. The minor variation trends  are however lost with this kind of treatment.}. The comparison of  Fig. \ref{fig:8} and Fig. \ref{fig:6} reveals that the slight non-linearity existed even before the weighting pre-treatment was performed. Most probably, this was caused by the effects of plural scattering in the spectra formation. Appendixes A and B prove that  non-linear mixing of compounds appears more likely  due to plural scattering while the weighting pre-treatment does not add noticeable extra distortion. Appendix B suggests that  dramatic deviation from the linear mixing model can be observed in cases when low-loss EEL spectra of constituting compounds are drastically different and TEM samples are thick. Such cases would require the application of non-linear mixing model  accounting for the interaction terms in spectra formation similar to those described in \cite{Keshava2002,Halimi2011,Altmann2013,Altmann2014,Halimi2017,Cavalcanti2019, Uezato2019}.

\subsection{Constraints for endmembers} \label{sec:4.1}

The majority of endmembers techniques put  certain constraints on the values of  endmember spectra and abundances in the course of their optimisation. These constraints should ensure the physical interpretability of the results. In contrast, the suggested method does not apply any explicit constraints to the positions of endmembers. This simplifies the algorithm and reduces the computational cost. Some  constraints are nevertheless satisfied implicitly. 

One of the common constraints is non-negativity of endmembers, which precludes the appearance of uninterpretable  spectra. We should stress, however, that this constraint is too weak for the case of EELS spectrum-images. EEL spectra appear usually on top of the significant background and are therefore situated far from the domain of negative values in factor space. On the other hand, unphysical behavior can manifest itself  in smooth deep wells instead of the expected peaks or in unlogical changes  of  the otherwise smoothly varying background. Such constraints are not easy to formulate and to implement into  algorithms. It is therefore strongly desired that  well interpretative endmembers evolve somehow naturally as a result of the algorithm. 

Although the original VCA approach  does not put explicitly constraints on the endmember spectra, the latter is fulfilled automatically as VCA chooses endmembers from the set of existing experimental data points and each of them has a physical meaning.   Strictly speaking, the endmembers extracted as described in Sections \ref{sec:3.4.1}, \ref{sec:3.4.2}  do not coincide with any of the existing data points. Nevertheless, they are always situated in the region of factor space densely occupied by existing data points. The algorithm averages over several endmembers falling into the same channel of the histogram (Section \ref{sec:3.4.1}).  It is possible that few data points from the same histogram channel  appear eventually in the drastically different branches of the data distribution and their averaging would  create an endmember in an unphysical domain of factor space. Such cases will be, however,  classified as outliers  in the forthcoming clustering step (Section \ref{sec:3.4.2}) and hence eliminated.

More generally, the extensive statistical filtering of the results by the procedures described in Sections \ref{sec:3.4.1}, \ref{sec:3.4.2} ensures that the calculated endmembers approach the positions of the true latent compounds and therefore satisfy all mentioned criteria. The eventual appearance of unphysical features in the endmember spectra or statistically significant negative values in the abundances would indicate that the procedure failed for some reason and that the certain algorithm parameters should be tuned. 

\subsection{Number of endmembers} \label{sec:4.2}

As described in Section  \ref{sec:3.4.2}, the algorithm allows to evaluate objectively the number of endmembers constituting a given dataset by computing the rating (statistical relevance) for each given endmember. The final choice is, however, reserved for the  user, who makes decisions based on the provided rating, the physical meaning and maybe some pre-knowledge. 

\begin{figure}[p]
\centering\includegraphics[width=0.5\linewidth]{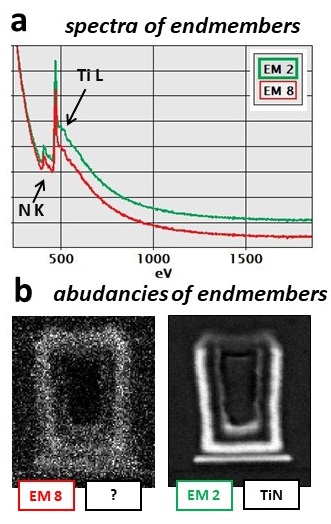}
\caption{(a) Spectrum  and (b) abundance map  of  endmember 8 extracted in Section \ref{sec:3.4.2} in comparison to  endmember 2 (TiN).}
\label{fig:9}
\end{figure}

In the considered example, we removed endmember 8, because its statistical relevance was questionable as seen from Table \ref{tab:2}. 
The evaluation of its spectrum reveals  proximity to the spectrum of  endmember 2, i.e. TiN (Fig. \ref{fig:9}a). Also the abundance map resembles closely the map of  endmember 2 while being strongly  smeared out spatially (Fig. \ref{fig:9}b). The comparison between taking into account 7 or 8 endmembers shows that the spectra and abundance maps  of the  endmembers 1-7 are almost the same for both variants. Therefore, endmember 8 can be safely discarded.  

The suggested algorithm has no built-in ability to evaluate whether a resulting endmember is physical or not. The endmember criteria are rather formal, namely endmembers must be situated at simplex corners in  factor space taking into account Gaussian smearing due to noise. These special positions have typically a clear physical meaning, however, the situation might be more complicated due to non-linearities in the spectra formation discussed in Section \ref{sec:4.0}.  In some cases,  the  variations between given compounds might noticeably deviate  from the shortest path, which could create  "pseudo-corners" in a multi-dimensional simplex and  increase the dimensionality of  factor space.  In the case of slight deviations from linearity, it is easier to deal with this problem by manual inspection of the  endmembers of smallest ratings.  The simplex pseudo-corners associated with the non-linearities are typically much weaker than those corresponding to real compounds and can be therefore easily identified. 

We can speculate more generally on the expected dimensionality of a given dataset. For instance, the considered example was known to consist of 11  compounds as determined by the combination of various analytical methods (Table \ref{tab:1}). Why the PCA analysis identified only 7 well-defined dimensions in the dataset? At the given experimental conditions for scanning and spectra acquisition, some compounds cannot be discriminated from each other. Such "washing-out" of minor compounds in the presence of insufficient statistical input was described in the literature \cite{Lichert2013, Potapov2017b,Jones2018}.  Except of the already mentioned  cases of TaN and HfO, the slightly different compositions TiN-A, TiN-B and TiN-C as well as SiO-A and SiO-B unfortunately cannot be resolved. 

\begin{figure}[ht]
\centering\includegraphics[width=0.5\linewidth]{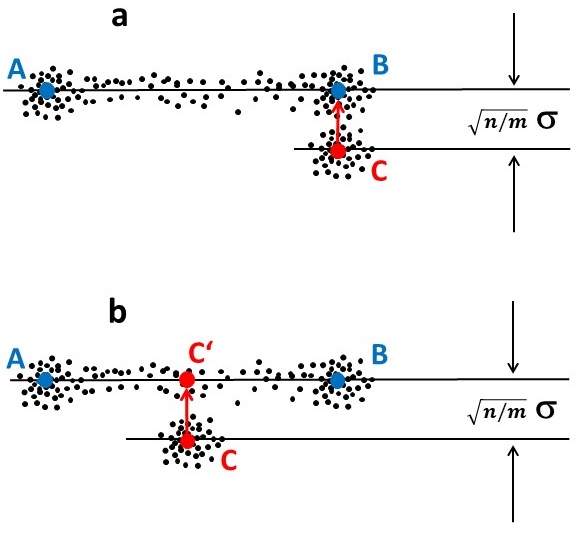}
\caption{Schematic illustration of (a)  merging two close compounds $\m B$ and $\m C$ in  factor space  and  (b) descending compound $\m C$  onto the line connecting $\m A$ and $\m B$.}
\label{fig:10}
\end{figure}

This can be understood from the schematic in Fig. \ref{fig:10}a. Hypothetical compounds $\m A$ and $\m B$ are well distinct while compound $\m C$ is situated quite closely to $\m B$ in factor space. Theoretical considerations of least-square based methods, like PCA, reveals that $\m C$ can be distinguished from $\m B$ only if they are separated by a distance greater than $\sigma \sqrt{\nicefrac{n}{m}}$ \cite{Nadler2008}. As previously, $m$ is the number of STEM probe positions, $n$ is the number of energy channels and $\sigma$ is the noise level. The dimensionality of  a dataset thus can appear lower than it might be expected. To resolve two compounds with close spectral proximity, one has to change the experimental conditions, namely to reduce the noise in the acquired spectra ($\sigma$ decreases) or increase the density of sampling ($m$ increases). This would improve the statistics and add additional dimensions to the PCA-reduced  factor space.

We finally consider the relationship between the dimensionality of PCA ($k$) and the number of endmembers ($r$). Simple geometrical arguments suggest that at least $(k+1)$ endmembers are required for the adequate description of data points in the $k$-dimensional space. The case  $r<(k+1)$ due to non-linearity in spectra formation was already discussed above. But could it happen that $r > (k+1)$? Strictly speaking, all distinct $r$ compounds would occupy well-defined points in factor space and therefore the condition $r = (k+1)$ should be fulfilled. However, in the presence of noise, the PCA dimensionality might be smaller than expected as explained above. Fig. \ref{fig:10}b shows such an example, where compound $\m C$ is located quite far from both $\m A$ and $\m B$ but close to the line connecting the $\m A$ and $\m B$ positions. In this situation, the statistical methods will determine a one-dimensional subspace and compound $\m C$ will be placed at the point $\m C'$ situated at the line connecting $\m A$ and $\m B$. Therefore, three endmembers could be identified in such a one-dimensional PCA space. As mentioned above, the probability of eventual merging the PCA dimensions increases with increasing  the noise level $ \sigma$ and with decreasing the number of STEM probe positions $m$ in a dataset.   

This analysis  suggests that the most probable number of endmembers is $r=(k+1)$ provided that the dimensionality $k$ of the PCA-reduced factor space was determined correctly. Nevertheless $r$ might deviate from $k+1$ in both direction due to the effects discussed above.         

\subsection{Applicability of the method} \label{sec:4.3}

For the sake of simplicity, the algorithm was described using only one example,  namely a typical characterization task in semiconductor industry. The performance of the algorithm was, however, tested at a number of real-life objects examined by both STEM EELS and EDX spectrum-imaging. As an example, Supplementary Material 5 shows the endmembers analysis of the same CMOS device  characterized by EDX. 

As outlined above, our algorithm combines  extensive geometrical manipulations with  Bayesian inference. This allows to profit from the advantages of both approaches. For the sake of comparison, we analysed the dataset described in Section \ref{sec:2} with other state-of-the-art Bayesian (Supplementary Material 3) and geometrical (Supplementary Material 4) methods. In both cases, our algorithm performed better in terms of accuracy of the extracted endmember spectra and maps. However, as EELS and EDX spectrum-images show quite variable internal structure, the actual performance might differ significantly from one to another dataset. More evaluation is needed to compare comprehensively the performance of these algorithms.

\section{Conclusions}

We have developed a new, simple and robust method  for extracting physically meaningful endmember spectra  and abundances from STEM spectrum-images. These endmembers  represent highly probably real latent compounds constituting a given dataset. The suggested method exhibits reasonable accuracy, high reproducibility and fast convergence. This method can be applied to a large scope of STEM EELS and EDX spectrum-imaging data. The DigitalMicrograph plugin for MSA utilized in the present work is available at http://temdm.com/web/msa/.

\section{Acknowledgement}

The authors appreciate support from ERC (grant 715620 under the Horizon 2020 program) and DFG project 431448015. The suggestions from an anonymous reviewer helped a lot in the rigorous formulation of the algorithm.

\section*{Appendix A: Non-linearity in the mixing model caused by weighting}

The weighting procedure (\ref{eq:2}) is a non-linear transformation that can potentially destroy linear mixing  among  endmembers. This appendix investigates how significantly the weighting procedure might distort the otherwise linear relationship. 

\begin{figure}[ht]
\centering\includegraphics[width=0.5\linewidth]{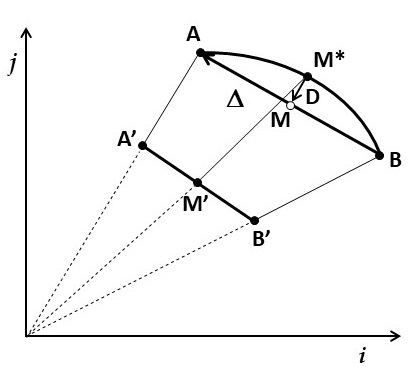}
\caption{Schematic action of the weighting procedure that transforms endmembers $\m A$ and $\m B$ into $\m A'$ and $\m B'$. The figure is drawn in abstract $i-j$ coordinates of factor space. The geometrical middle points between $\m A$ and $\m B$ and $\m A'$ and $\m B'$ are denoted as $\m M$ and $\m M'$,  respectively.   As a measure of non-linearity we take the difference $\m D$ between points $\m M$ and    $\m M^*$ that is obtained from $\m M'$ by the un-weighting transformation. }
\label{fig:11}
\end{figure}

Consider for brevity only two endmembers $\m A$ and $\m B$ that show a linear mixture along the line $\m{A-B}$ in factor space (Fig. \ref{fig:11}). Both $\m A$ and $\m B$ are $n$-dimensional vectors representing spectra composed of $n$ energy channels. After the weighting procedure they become respectively $\m A'$ and $\m B'$ connected by the non-straight curve $\m {A'\sim B'}$. Equivalently, the straight line $\m A'-\m B'$ becomes curved  $\m {A\sim B}$ after the un-weighting procedure. Lets define a middle point $\m M$ with the equal contributions from $\m A$ and $\m B$ and a corresponding point $\m M'$ in between of $\m A'$ and $\m B'$. As a measure for non-linearity, we will take the distance $\m D$ between  $\m M$ and $\m M^*$, which is a point obtained from $\m M'$ by applying the un-weighting procedure (see Fig.\ref{fig:11}). 

To simplify the evaluation, we consider a simple weighting procedure where each  coordinate of a vector $\m X$ is replaced by its square root: $x_i \rightarrow \sqrt{x_i}$. This corresponds closely to the Anscombe transform when $x_i \gg 0$, as in the case of EEL spectra. Then
\begin{equation}
|\m D|^2 = \sum_{i}^{}\left( \frac{a_i+b_i}{2} -\left( \frac{\sqrt{a_i}+\sqrt{b_i}}{2} \right)^2 \right)^2
\label{eq:EB1}
\end{equation}
\noindent where $a_i$ and $b_i$ are coordinates of $\m A$ and $\m B$, respectively. We define also a vector $\Delta =\m A - \m B$. In terms of $\m M$ and $\Delta$, (\ref{eq:EB1}) is recast as
\begin{equation}
|\m D|^2 = \sum_{i}^{}\left( m_i -\left( \frac{\sqrt{m_i-\nicefrac{\delta_i}{2}}+\sqrt{m_i+\nicefrac{\delta_i}{2}}}{2} \right)^2 \right)^2
\label{eq:EB2}
\end{equation}
\noindent where $m_i$ and $\delta_i$ are coordinates of $\m M$ and $\Delta$ respectively. Expanding the inner squared brackets yields:
\begin{equation}
|\m D|^2 = \sum_{i}^{}\left(\frac{m_i}{2} - \frac{\sqrt{m_i-\nicefrac{\delta_i}{2}}\sqrt{m_i+\nicefrac{\delta_i}{2}}}{2}  \right)^2 = \sum_{i}^{}\left(\frac{m_i}{2} - \frac{\sqrt{m_i^2-\nicefrac{\delta_i^2}{4}}}{2}  \right)^2
\label{eq:EB3}
\end{equation}
After expanding the outer squared brackets, (\ref{eq:EB3}) becomes:
\begin{equation}
|\m D|^2 = \sum_{i}^{}\left( \frac{m_i^2}{2} - \frac{\delta_i^2}{16} - \frac{m_i \sqrt{m_i^2-\nicefrac{\delta_i^2}{4}}}{2}   \right)
\label{eq:EB4}
\end{equation}
The expression under the square root can be decomposed in the Taylor series: $\sqrt{x^2+d} = x(1+\nicefrac{d}{2x^2}-\nicefrac{d^2}{8x^4}+...)$. Limiting the Taylor series decomposition to the quadratic term, expression (\ref{eq:EB4})  simplifies as:
\begin{equation}
|\m D|^2 = \sum_{i}^{} \left( \frac{m_i^2}{2} - \frac{\delta_i^2}{16} -\frac{m_i^2}{2}(1-\frac{\delta_i^2}{8m_i^2}-\frac{\delta_i^4}{128m_i^4}) \right) =   \frac{1}{256}\sum_{i}^{} \delta_i^2 \left(\frac{\delta_i}{m_i} \right)^2
\label{eq:EB5}
\end{equation}
Taking the squared \textit{relative} difference finally gives:

\begin{equation}
\frac{|\m D|^2}{|\Delta|^2}   = \frac{1}{256} \frac{\sum_{i}^{} \delta_i^2 \left(\frac{\delta_i}{m_i} \right)^2}{\sum_{i}^{} \delta_i^2 }
\label{eq:EB6}
\end{equation}
The sum in the  nominator consists of the weighting coefficients $\nicefrac{\delta_i}{m_i}$. Eventually, few  $m_i$ might be zero, which would cause the divergence of (\ref{eq:EB6}). This is, however, solely the consequence of ignoring the higher order terms in the Taylor series decomposition (\ref{eq:EB5}). Note that $|\m D|$ does not depend on the choice of the coordinates. It is always possible to choose a system of coordinates such that none of the $m_i$s are zero. Furthermore,  $|\m M|$ is typically larger than $|\Delta|$ in  EDX,  while it is always much larger than $|\Delta|$ in EELS. If  both $\m M$ and $\Delta$ are smooth spectra, the fractions $\nicefrac{\delta_i}{m_i}$ on the right side of (\ref{eq:EB6}) are expected to be  less than 1. Therefore

\begin{equation}
\frac{|\m D|^2}{|\Delta|^2}   < \frac{1}{256} 
\label{eq:EB7}
\end{equation}

\noindent and

\begin{equation}
\frac{|\m D|}{|\Delta|}    < \frac{1}{16} 
\label{eq:EB8}
\end{equation}

\section*{Appendix B: Non-linearity in the mixing model caused by plural scattering}

The complexity of the EEL spectra formation can distort the otherwise linear mixing among the endmembers. This appendix  estimates the non-linearity induced by plural scattering of  detected electrons.

\begin{figure}[ht]
\centering\includegraphics[width=1.0\linewidth]{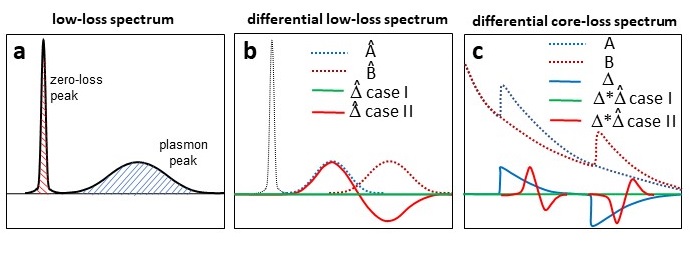}
\caption{(a) Schematic shape of a low-loss EEL spectrum consisting of a zero-loss and a plasmon peak. The area under the plasmon peak is 0.3-1.0 of that for the zero-loss peak (not shown in a realistic intensity scale). (b) Shapes of differential low-loss spectra for case I (green), when two plasmon peaks are similar in position and  width  and for case II (red), when two plasmon peaks are drastically different. (c) Differential core-lose spectrum $\Delta$ (blue) and an oscillatory curve obtained by the convolution of $\Delta$ and $\hat{\Delta}$ for case I (green) and II (red).}
\label{fig:12}
\end{figure}

As in Appendix A, consider $n$-dimensional vectors representing spectra composed of $n$ energy channels. The effect of plural scattering on core-loss EEL spectra  can be described as a convolution of a (single scattering) core loss spectrum with a normalized low-loss spectrum ((4.35) in \cite{Egerton}):

\begin{equation}
\m J = \m S  * \left(\m I +\m L \right)
\label{eq:EC0}
\end{equation}

\noindent 
where $\m J$ is the total spectrum, $\m S$ is a single scattering core-loss  spectrum, $\m L$ is the plural scattering  generated in the low-loss energy region normalized to the total number of the incoming electrons and $\m I$ is a discrete analog of the delta-function, i.e., the vector consisting of unity at the first row and zeros at the other rows. Formula (4.35) from \cite{Egerton} is rewritten here in the discrete form where the convolution of two vectors $\m X$ and $\m Y$ is defined through $\sum_{j}^{}X(j)Y(i-j)$.  The most prominent part of $\m L$ is  a plasmon peak as schematically shown in  Fig. \ref{fig:12}a.  The position and the width of the plasmon peak can vary significantly even for chemically similar compounds (e.g. plasmon peak is observed at ~16eV in $Si$ and at ~24eV in $SiO_2$) while the total area under the plasmon peak is typically 0.3-1.0 of that pertaing to the zero-loss peak. That roughly corresponds to the typical thickness of TEM samples expressed  in terms of the effective inelastic mean-free path length. 

As in Appendix A, we consider two endmembers  with unique single-scattering core-loss spectra $\m A$ and $\m B$ that  mix linearly in a given object. Plural inelastic scattering can, however,  destroy the linear relationship between $\m A$ and $\m B$ in factor space. In the presence of plural scattering, the endmember spectra become $\m A * (\m I+\hat{\m A})$ and $\m B * (\m I+\hat{\m B})$ where $\hat{\m A}$ and $\hat{\m B}$ are  normalized low-loss spectra of the corresponding endmembers. 

Similar to Appendix A, we consider a 50:50 mixture of the two endmembers and compare the resulting spectrum with the geometric midpoint in factor space:  

\begin{equation}
\m M =  \frac{\m A +\m B}{2} * (\m I+\frac{\hat{\m A} +\hat {\m B}}{2}) 
\label{eq:EC1}
\end{equation}

\begin{equation}
\m M' =  \frac{\m A * (\m I+\hat{\m A}) +\m B * (\m I+\hat{\m B})}{2} 
\label{eq:EC2}
\end{equation}

\noindent
where (\ref{eq:EC1}) is the actual spectrum of the 50:50 mixture and (\ref{eq:EC2}) is the geometrical mean between $\m A$ and $\m B$.

Performing the discrete Fourier transform of (\ref{eq:EC1}) yields
\begin{equation}
\m M^F =  \left( \frac{\m A +\m B}{2}\right)^F \left(1 + \frac{(\hat{\m A} +\hat {\m B})}{2}\right)^F
\label{eq:EC3}
\end{equation}

\noindent
where the superscript $F$ denotes the Fourier transform of the corresponding vector.

Introducing the differential core-loss $\Delta = \m A - \m B$ and low-loss  $\hat{\Delta} = \hat{\m A} - \hat{\m B}$ spectra and taking into account the linearity of the Fourier transformation, (\ref{eq:EC3}) can be rewritten as:
\begin{equation}
\m M^F =  \frac{\m A^F (1+ \hat{\m A}^F) +\m B^F (1+ \hat {\m B}^F)}{2} - \frac{\Delta^F \hat{\Delta}^F}{4} 
\label{eq:EC4}
\end{equation}
Performing the inverse Fourier transformation then gives

\begin{equation}
\m M =  \frac{\m A * (\m I + \hat{\m A}) +\m B * (\m I+\hat {\m B})}{2} - \frac{\Delta * \hat{\Delta}}{4} 
\label{eq:EC5}
\end{equation}
Therefore the difference in the position of the 50:50 mixture and its geometric middle point in factor space is

\begin{equation}
\m D = \m M -\m M' =  - \frac{\Delta * \hat{\Delta}}{4} 
\label{eq:EC6}
\end{equation}

\noindent and the relative difference is
\begin{equation}
\frac{|\m D|^2}{|\Delta|^2}   = \frac{1}{16} \frac{\sum_{i}^{} \delta_i^{*2}}{\sum_{i}^{} \delta_i^2}
\label{eq:EC7}
\end{equation}

\noindent where $\delta_i^*$ are coordinates of  $\Delta * \hat{\Delta}$. In other words, they represent an energy distribution composed of the differential core-loss spectrum  convoluted with the differential low-loss spectrum. 

To understand the meaning of $\Delta * \hat{\Delta}$ we consider two extreme cases:  (case I) when the plasmon peak positions and widths for both compounds are almost identical and (case II) when they are drastically different. Provided that the sample thickness is uniform along the investigated area, the differential spectrum $\hat{\Delta}$ is nearly zero in case I and represents an oscillatory curve like that in Fig. \ref{fig:12}b in case II.   The differential core-loss spectrum $\Delta$ is shown schematically in Fig. \ref{fig:12}c. In case I, the convolution of $\Delta$ with zero-valued $\hat{\Delta}$ would give zero (green curve in Fig. \ref{fig:12}c). In case II, this would be, however, a complicated oscillatory curve shown as red line in Fig. \ref{fig:12}c. It is quite difficult to estimate  (\ref{eq:EC7}) in this case. At the worst, the nominators and denominators in (\ref{eq:EC7}) may be of comparable size. Therefore, we can constrain the relative difference only as

\begin{equation}
\frac{|\m D|^2}{|\Delta|^2}   < \frac{1}{16} \, (\text{for \, relative \, thickness} <1)
\label{eq:EC8}
\end{equation}

Note that  estimation (\ref{eq:EC8}) is valid for the thickness of TEM samples less than 1 (in terms of the effective inelastic mean-free path). When increasing the sample thickness above this limit, the  deviation from linearity might rise dramatically. 

\section*{Appendix C: Influence of the interior data distribution on the shape of the noisy tail}

In Section \ref{sec:3.4.1}, it was assumed that the left tail of the observed histogram distribution $X(i)$ follows approximately the shape of the left tail of the sampled Gaussian kernel. Strictly speaking, this is only valid if the interior histogram channels to the right of  endmember channel $c$  are not populated.  For brevity, we will call this a \textit{single-channel} approximation.  In reality,  the interior noise-free data distribution $Y(i)$ will also influence the expected shape of the tail. This appendix evaluates the validity of the \textit{single-channel} approximation against the realistic \textit{multi-channel} signals.

Consider the range of histogram channels $[c-d,c]$, where $d$ is the truncation limit for the sampled Gaussian kernel $S$. In the single-channel approximation, the observed counts are expected to approach the shape of kernel $S$:

\begin{equation}
X(i) \approx S(i) = \frac{\alpha(c)}{\sqrt{2\pi \sigma^*}} \exp(- \frac{(c-i)^2} {2 \sigma^{*2}})
\label{eq:C1}
\end{equation}

\noindent where $\alpha (c)$ is the "true", noise free count at channel $c$. Formula (\ref{eq:C1}) implies large intensity limit where the finite number of counts does not play a role.

In the plural-channel picture, there is a continuous noise-free distribution $Y(i)$ to the right from endmember channel $c$. Our task is to evaluate how the sharp edge of noise-free distribution $Y(i)$ will be changed in the presence of Gaussian smearing. The observed counts should approach a kernel $\m M$, which is a convolution of $Y$ and $S$:

\begin{equation}
X(i) \approx M(i) = \sum_{j=0}^{d}\frac{\alpha(c+j)}{\sqrt{2\pi \sigma^*}} \exp(- \frac{(c+j-i)^2} {2 \sigma^{*2}})
\label{eq:C2}
\end{equation}

\noindent 
where we replaced the noise-free counts $Y(i)$ for $\alpha(c+j)$ in order to be consistent with the notation in (\ref{eq:C1}). Similar to (\ref{eq:C1}), (\ref{eq:C2}) is  valid for the histogram range $[c-d,c]$.

Expanding the squared brackets and moving the $j$-independent term out of the sum yields

\begin{equation}
M(i) =  \exp(- \frac{(c-i)^2}{2 \sigma^{*2}}) \sum_{j=0}^{d}\left( \frac{\alpha(c+j)}{\sqrt{2\pi \sigma^*}} \exp(- \frac{j^2}{2 \sigma^{*2}})\exp(- \frac{j(c-i)}{\sigma^{*2}}) \right)
\label{eq:C3}
\end{equation}
We replace $M(i)$  for

\begin{equation}
\tilde{S}(i) = \frac{\alpha'}{\sqrt{2\pi \sigma^*}} \exp(- \frac{(c-i)^2} {2 \sigma^{*2}})
\label{eq:C4}
\end{equation}
where 

\begin{equation}
\tilde{\alpha} = \sum_{j=0}^{d} \alpha (c+j) \exp(- \frac{j^2}{2 \sigma^{*2}}) 
\label{eq:C5}
\end{equation}
Kernel $\tilde{S}(i)$ is just a rescaled version of $S(i)$, where $\alpha(c)$ is replaced for effective $\tilde{\alpha}$. The deviation of the multi-channels picture (\ref{eq:C3}) from the single-channel approximation (\ref{eq:C4}) is 

\begin{equation}
\frac{M(i)-\tilde{S}(i)}{\tilde{S}(i)}= \frac{\sum_{j=0}^{d} \alpha(c+j) \exp(- \frac{j^2}{2 \sigma^{*2}}) (\exp(- \frac{j(c-i)}{ \sigma^{*2}}) -1)}{\sum_{j=0}^{d} \alpha(c+j) \exp(- \frac{j^2}{2 \sigma^{*2}})} 
\label{eq:C6}
\end{equation}

\begin{figure}[ht]
\centering\includegraphics[width=0.8\linewidth]{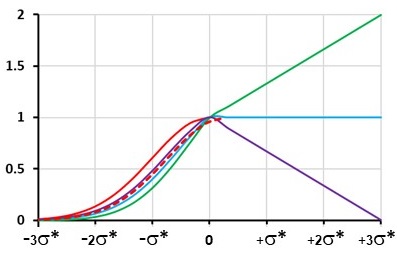}
\caption{Left tail of the sampled Gaussian kernel (red) centered at endmember channel $c$ and shifted at $0.25\sigma^*$ from $c$ (red dashed). The other curves show the expected smoothing of the sharp distribution edge if the "true" distribution is constant (blue), sharply increases (green) or sharply drops (violet). The abscissa is scaled in the units of the noise standard deviation $\sigma ^*$ and the coordinate origin is chosen at $c$. For the better comparison, all distributions are scaled such as they equal 1 at $c$.}
\label{fig:13}
\end{figure}

Fig. \ref{fig:13} shows simulated distribution tails assuming several models for the interior data distributions $Y(i)$ - the homogeneous distribution and sharp linear rise or decrease of counts. In all cases, the general shape of the distribution tail is approximately conserved although Fig.\ref{fig:13} suggests that the single-channel approximation tends to shift slightly (0.2 - 0.5$\sigma$) the position of the "true" endmember channel $c$. 

\section*{Appendix D: Scaling the noisy tail of data distribution} 

Section \ref{sec:3.4.1} evaluates the likelihood of the hypothesis that the endmember is situated at a certain channel $c$ of the observed data histogram $X(i)$.  In that case, all  counts at the channels to the left from $c$ arise from Gaussian noise and should follow  the left tail of the sampled Gaussian kernel as in Equation (\ref{eq:C1}). The scaling parameter $\alpha(c)$  is, however, unknown. This appendix establishes the choice of $\alpha(c)$  that maximizes the likelihood in (\ref{eq:9}).

The likelihood that the count at a given channel $i$ satisfies the hypothesis  $C=c$ is

\begin{equation}
\mathcal{P}(X_i|C=c) =  \frac{1}{\sqrt{2\pi \alpha(c) S(c-i)}} \exp{\left(- \frac{(X(i)-\alpha(c) S(c-i))^2}{2\alpha(c) S(c-i)}\right)}
\label{eq:EA1}
\end{equation}

\noindent
where $S(j)$ is a sampled Gaussian kernel centered at channel $c$  and $\alpha$ is the scaling parameter as depicted in Fig. \ref{fig:5}a. Equation (\ref{eq:EA1}) implies the Poisson statistics in the deviation of the observed counts from the true distribution. The likelihood that \textit{all} the channels to the left from $c$ satisfy the hypothesis $C=c$ is
 
\begin{equation}
\mathcal{P}(X_{c-d},...X_c|C=c) =
\prod_{i=c-d}^{c} \frac{1}{\sqrt{2\pi \alpha(c) S(c-i)}} \exp{\left (- \frac{(X(i)-\alpha(c) S(c-i))^2}{2\alpha(c) S(c-i)}\right)}
\label{eq:EA2}
\end{equation}

\noindent
where the multiplication of the probabilities is carried out  within a certain range of channels $[c-d,c]$. The choice of parameter $\alpha$ should maximize the total likelihood (\ref{eq:EA2}). In the logarithmic scale, Equation (\ref{eq:EA2}) becomes

\begin{equation}
\ln{\mathcal{P}(X_{c-d},...X_c|C=c)} = \sum_{i=c-d}^{c}\left(-\frac{1}{2}(\ln({2\pi S(c-i)}) + \ln{\alpha(c)})- \frac{(X(i)-\alpha(c) S(c-i))^2}{2\alpha(c) S(c-i)} \right)
\label{eq:EA3}
\end{equation}

After some algebraic transformations (\ref{eq:EA3}) becomes:

\begin{equation}
\begin{split}
& \ln{\mathcal{P}(X_{c-d},...X_c|C=c)} =\\&  
\sum_{i=c-d}^{c}\left(-\frac{1}{2}(\ln({2\pi S(c-i)}) + \ln{\alpha(c)})- \frac{X^2(i)}{2\alpha(c) S(c-i)} -\frac{\alpha(c)S(c-i)}{2} +X(i) \right)
\end{split}
\label{eq:EA4}
\end{equation}

Taking the derivative of (\ref{eq:EA4}) with respect to $\alpha(c)$ and equating the result with zero, we obtain: 

\begin{equation}
\sum_{i=c-d}^{c}\left(-\frac{1}{\alpha(c)}+ \frac{X^2(i)}{\alpha^2(c) S(c-i)} -S(c-i) \right) = 0
\label{eq:EA5}
\end{equation}
In case of sufficient counts in the histogram channels, $\alpha$ is expected to be noticeably higher than 1 while the remaining terms are close to unity, therefore the first term in Equation (\ref{eq:EA5}) can be neglected. Rearranging slightly the second term yields

\begin{equation}
\sum_{i=c-d}^{c}\left( \frac{X(i)}{\alpha(c)} \frac{X(i)}{\alpha(c) S(c-i)}- S(c-i)\right) \approx{} 0
\label{eq:EA7}
\end{equation}

\noindent
To determine $\alpha(c)$ we write a recurrent formula:

\begin{equation}
\alpha (c) \approx{} \sum_{i=c-d}^{c}X_i \frac{X(i)}{\alpha(c)S(c-i)} / \sum_{i=c-d}^{c}S(c-i)
\label{eq:EA8}
\end{equation}

\noindent
where the terms $\frac{X(i)}{\alpha(c)S(c-i)}$ are close to unity. In the first approximation $\alpha (c)$ is

\begin{equation}
\alpha_1 (c) = \sum_{i=c-d}^{c}X(i) / \sum_{i=c-d}^{c}S(c-i)
\label{eq:EA9}
\end{equation}

\noindent
The next approximations can be found as

\begin{equation}
\alpha_n (c) = \sum_{i=c-d}^{c}X_i\frac{X(i)}{\alpha_{n-1}(c) S(c-i)} / \sum_{i=c-d}^{c}S(c-i)
\label{eq:EA10}
\end{equation}
As we are searching for an approximate solution of (\ref{eq:9}), the first approximation for $\alpha(c)$ is sufficient. 

\section*{Appendix E: Alternative prior distributions} 

In few cases, the Bayesian inference with the equal prior distribution (\ref{eq:8}) introduced in Section \ref{sec:3.4.1} generates outliers. This happens when the shape of the histogram somewhere in its middle  eventually mimics the shape of a Gaussian kernel. To suppress this unwanted behaviour we consider alternative priors for the left-side endmember position that satisfy the conditions:

\begin{equation}
\mathcal{P}[C=c] 
\begin{cases}
    \text{nearly equal at the left side of histogram} \\
    \text{rapidly decreases in the middle of histogram}
\end{cases}
\label{eq:E1}
\end{equation}

The simplest option is a prior quadratically decaying in the range $[0,...,l]$  
\begin{equation}
\mathcal{P}[C=c] = \frac{(1-(\frac{i}{l})^2)}{\sum_{i=1}^{l}(1-(\frac{i}{l})^2)}
\label{eq:E2}
\end{equation}

\noindent shown in pink in Fig. \ref{fig:5}b. This is however a rather subjective prior reflecting only our believe in (\ref{eq:E1}). 

We will construct a less subjective prior that satisfies (\ref{eq:E1}) but utilises some information from  counts $X(i)$ observable in the histogram range $[0,l]$. As a starting point we assume no knowledge about the position of endmember $c$. We also know that the "true" noise-free data distribution is smeared out by a certain smearing kernel $K$ due to noise. In contrast to Section \ref{sec:3.4.1} and Appendixes C,D, we do not make any strong assumptions about $K$. We only assume that $K$ is symmetric, i.e. its left and right tails are mirrored. Then $\mathcal{P}[C=l] =0$ as all observable counts $X(i)$ are at the left from $c$, which contradicts the assumption of symmetrical $K(j)$. In general, more counts $X(i)$ are observed at the left from $c$, more it constraints the possible width and shape of kernel $K(j)$, and therefore the hypothesis $C=c$ is less probable. Accordingly, we construct the prior as

\begin{equation}
\mathcal{P}[C=c] = \frac{1-\frac{1}{S}\sum_{i=1}^{c}X(i)}{\sum_{c=1}^{l}(1-\frac{1}{S}\sum_{i=1}^{c}X(i))}
\label{eq:E3}
\end{equation}

\noindent where $S = \sum_{i=1}^{l} X(i)$ is the total counts in the histogram.

The argumentation above is a simplest form of the empirical Bayesian method for constructing priors. This allows to estimate very roughly the general shape (not point-to-point variation) of the probability distribution based on  given $X(i)$. Prior (\ref{eq:E3}) (shown in green in Fig. \ref{fig:5}b) was actually used in the current version of the code.

\bibliography{MSA.bib}

\begin{thebibliography}{57}
\providecommand{\natexlab}[1]{#1}
\providecommand{\url}[1]{\texttt{#1}}
\expandafter\ifx\csname urlstyle\endcsname\relax
  \providecommand{\doi}[1]{doi: #1}\else
  \providecommand{\doi}{doi: \begingroup \urlstyle{rm}\Url}\fi

\bibitem[Pearson(1901)]{Pearson1901}
K.~Pearson.
\newblock On lines and planes of closest fit to systems of points in space.
\newblock \emph{Philosophical Mag.}, 2:\penalty0 559--572, 1901.

\bibitem[Hotelling(1933)]{Hotelling1933}
H.~Hotelling.
\newblock Analysis of a complex of statistical variables into principal
  components.
\newblock \emph{Journal of Educational Psycholog}, 24\penalty0 (6):\penalty0
  417–441, 1933.

\bibitem[Tipping and Bishop(1997)]{Tipping1997}
M.~E. Tipping and C.~M. Bishop.
\newblock Probabilistic principal component analysis.
\newblock \emph{J. Royal Statist. Soc.B}, 61:\penalty0 611--622, 1997.

\bibitem[Jolliffe(2002)]{Jolliffe2002}
I.~T. Jolliffe.
\newblock \emph{Principal Component Analysis}.
\newblock Springer Verlag, 2nd edition, 2002.

\bibitem[Malinowski(2002)]{Malinowski2002}
E.~R. Malinowski.
\newblock \emph{Factor analysis in Chemistry}.
\newblock Wiley, 3rd edition, 2002.

\bibitem[Jolliffe and Cadima(2016)]{Jolliffe2016}
I.~T. Jolliffe and J.~Cadima.
\newblock Principal component analysis: a review and recent developments.
\newblock \emph{Philosophical Trans. A}, 374:\penalty0 20150202, 2016.

\bibitem[Comon(1994)]{Comon1994}
P.~Comon.
\newblock Independent component analysis, a new concept?
\newblock \emph{Signal Processing}, 36\penalty0 (287-314), 1994.

\bibitem[Hyvärinen(2001)]{Hyvarinen2001}
A.~Hyvärinen.
\newblock \emph{Independent component analysis}.
\newblock NY: Wiley, 2001.

\bibitem[Hyvärinen and Oja(2013)]{Hyvarinen2013}
A.~Hyvärinen and E.~Oja.
\newblock Independent component analysis: recent advances.
\newblock \emph{Philosophical Transactions: Mathematical, Physical and
  Engineering Sciences}, 371:\penalty0 20110534, 2013.

\bibitem[Tarantola and Valette(1982)]{Tarantola1982}
A.~Tarantola and B.~Valette.
\newblock Generalized nonlinear inverse problems solved using the least squares
  criterion.
\newblock \emph{Reviews of Geophysics and Space Physics}, 20:\penalty0
  219--232, 1982.

\bibitem[Plaza et~al.(2004)Plaza, Martínez, Pérez, and Plaza]{Plaza2004}
A.~Plaza, P.~Martínez, R.M. Pérez, and J.~Plaza.
\newblock A quantitative and comparative analysis of endmember extraction
  algorithms from hyperspectral data.
\newblock \emph{{IEEE} Transactions on Geoscience and Remote Sensing},
  42\penalty0 (3):\penalty0 650--663, 2004.

\bibitem[Martinez et~al.(2006)Martinez, Pérez, Plaza, Aguilar, Cantero, and
  Plaza]{Martinez2006}
J.~Martinez, R.M. Pérez, A.~Plaza, P.L. Aguilar, M.~Cantero, and J.~Plaza.
\newblock Endmember extraction algorithms from hyperstructural images.
\newblock \emph{Annals of Geophysics}, 49\penalty0 (1):\penalty0 93--101, Dec
  2006.

\bibitem[Bioucas-Dias et~al.(2012)Bioucas-Dias, Plaza, Dobigeon, Parente, Du,
  Gader, and Chanussot]{Bioucas-Dias2012}
J.M. Bioucas-Dias, A.~Plaza, N.~Dobigeon, M.~Parente, Q.~Du, P.~Gader, and
  J.~Chanussot.
\newblock Hyperspectral unmixing overview: geometrical, statistical, and sparse
  regression-based approaches.
\newblock \emph{IEEE Journal of selected topics in applied earth observations
  and remote sensing}, 5\penalty0 (2):\penalty0 354--379, 2012.

\bibitem[Tauler(1995)]{Tauler1995}
R.~Tauler.
\newblock Multivariate curve resolution applied to second order data.
\newblock \emph{Chemometrics and Intelligent Laboratory Systems}, 30:\penalty0
  133--146, 1995.

\bibitem[Ruckebusch and Blanchet(2013)]{Ruckebusch2013}
C.~Ruckebusch and L.~Blanchet.
\newblock Multivariate curve resolution: A review of advanced and tailored
  applications and challenges.
\newblock \emph{Analytica Chimica Acta}, 765:\penalty0 28--36, 2013.

\bibitem[Lavoie et~al.(2016)Lavoie, Braidy, and Gosselin]{Lavoie2016}
F.~B. Lavoie, N.~Braidy, and R.~Gosselin.
\newblock Including noise characteristics in {MCR} to improve mapping and
  component extraction from spectral images.
\newblock \emph{Chemom. Intell. Lab}, 153:\penalty0 40--50, 2016.

\bibitem[Potapov(2016)]{Potapov2016}
P.~Potapov.
\newblock Why principal component analysis of {STEM} spectrum images results in
  abstract, uninterpretable loadings?
\newblock \emph{Ultramicroscopy}, 160:\penalty0 197--212, 2016.

\bibitem[Kotula et~al.(2003)Kotula, Keenan, and Michael]{Kotula2003}
P.G. Kotula, M.R. Keenan, and J.R. Michael.
\newblock Automated analysis of {EDS} spectrum images in a {SEM}: a powerful
  new microanalysis technique.
\newblock \emph{Microsc. Microanal.}, 9:\penalty0 1--17, 2003.

\bibitem[Keenan(2009)]{Keenan2009}
M.R. Keenan.
\newblock Exploiting spatial-domain simplicity in spectral image analysis.
\newblock \emph{Surf. Interface Anal.}, 41:\penalty0 79--87, 2009.

\bibitem[Lucas et~al.(2013)Lucas, Burdet, Cantoni, and Hebert]{Lucas2013}
G.~Lucas, P.~Burdet, M.~Cantoni, and C.~Hebert.
\newblock Multivariate statistical analysis as a tool for the segmentation of
  {3D} spectral data.
\newblock \emph{Micron}, 52-53:\penalty0 49--56, 2013.

\bibitem[Winter(1999)]{Winter1999}
M.E. Winter.
\newblock N-fndr: An algorithm for fast autonomous endmember determination in
  hyperspectral data.
\newblock \emph{Proc SPIE image spectroscopy}, 3753:\penalty0 266--277, 1999.

\bibitem[Nascimento(2005{\natexlab{a}})]{Nascimento2005a}
M.P. Nascimento.
\newblock Vertex component analysis: a fast algorithm to unmix hyperspectral
  data.
\newblock \emph{IEEE Transactions on Geoscience and remote sensing},
  43\penalty0 (4):\penalty0 898--910, 2005{\natexlab{a}}.

\bibitem[Craig(1994)]{Craig1994}
M.~Craig.
\newblock Minimum volume transforms for remotely sensed data.
\newblock \emph{IEEE Transactions on geoscience and remote sensing},
  32\penalty0 (3):\penalty0 542--552, 1994.

\bibitem[Moussaoui et~al.(2006)Moussaoui, Carteret, Brie, and
  Mohhamad-Jafari]{Moussaoui2006}
S.~Moussaoui, C.~Carteret, D.~Brie, and A.~Mohhamad-Jafari.
\newblock Bayesian analysis of spectral mixture data using {M}arkov chains
  {M}onte-{C}arlo methods.
\newblock \emph{Chemometrics and Intelligent Laboratory Systems}, 81\penalty0
  (2):\penalty0 137--148, 2006.

\bibitem[Dobigeon et~al.(2009)Dobigeon, Moussaoui, Coulon, Tourneret, and
  Hero]{Dobigeon2009}
N.~Dobigeon, S.~Moussaoui, M.~Coulon, J.-Y. Tourneret, and A.O. Hero.
\newblock Joint {B}ayesian endmember extraction and linear unmixing for
  hyperspectral imagery.
\newblock \emph{IEEE Transactions on Signal Processing}, 57\penalty0
  (11):\penalty0 4355--4368, 2009.

\bibitem[Arngren et~al.(2011)Arngren, Schmidt, and Larsen]{Arngren2011}
M.~Arngren, M.N. Schmidt, and J.~Larsen.
\newblock Unmixing of hyperspectral images using {B}ayesian nonnegative matrix
  factorization with volume prior.
\newblock \emph{Journal of Signal Processing Systems}, 65:\penalty0 479–496,
  2011.

\bibitem[Dobigeon and Brun(2012)]{Dobigeon2012}
N.~Dobigeon and N.~Brun.
\newblock Spectral mixture analysis of eels spectrum-images.
\newblock \emph{Ultramicroscopy}, 120:\penalty0 25--34, 2012.

\bibitem[Shiga et~al.(2016)Shiga, Tatsumi, Muto, Tsuda, Yamamoto, Mori, and
  Tanji]{Shiga2016}
M.~Shiga, K.~Tatsumi, S.~Muto, K.~Tsuda, Y.~Yamamoto, T.~Mori, and T.~Tanji.
\newblock Sparse modelling of {EELS} and {EDX} spectral imaging data by
  nonnegative matrix factorization.
\newblock \emph{Ultramicroscopy}, 170:\penalty0 43--59, 2016.

\bibitem[Braidy and Gosselin(2019)]{Braidy2019}
N.~Braidy and R.~Gosselin.
\newblock Unmixing noisy co-registered spectrum images of multicomponent
  nanostructures.
\newblock \emph{Nature Scientific Reports}, 9:\penalty0 18797, 2019.

\bibitem[Paatero and Tapper(1994)]{Paatero1994}
P.~Paatero and U.~Tapper.
\newblock Positive matrix factorization: A non-negative factor model with
  optimal utilization of error estimates of data values.
\newblock \emph{Environmetrics}, 5\penalty0 (2):\penalty0 111–126, 1994.

\bibitem[Nascimento(2005{\natexlab{b}})]{Nascimento2005b}
M.P. Nascimento.
\newblock Does independent component analysis play a role in unmixing
  hyperspectral data ?
\newblock \emph{IEEE Transactions on geoscience and remote sensing},
  43\penalty0 (1):\penalty0 175--187, 2005{\natexlab{b}}.

\bibitem[Bonnet and Nuzollard(2005)]{Bonnet2005}
N.~Bonnet and D.~Nuzollard.
\newblock Independent components analysis: a new possibility for analysis a
  series of electron energy loss spectra.
\newblock \emph{Ultramicroscopy}, 102:\penalty0 327--337, 2005.

\bibitem[Yamazaki et~al.(2011)Yamazaki, Kotaka, and Kataoka]{Yamazaki2011}
T.~Yamazaki, Y.~Kotaka, and Y.~Kataoka.
\newblock Analysis of {EEL} spectrum of low-loss region using the
  {Cs}-corrected {STEM–EELS} method and multivariate analysis.
\newblock \emph{Ultramicroscopy}, 111\penalty0 (5):\penalty0 303--308, 2011.

\bibitem[Cattell(1994)]{Cattel1964}
R.B. Cattell.
\newblock The scree test for the number of factors.
\newblock \emph{Multivariate Behavioral Research}, 1\penalty0 (2):\penalty0
  245–276, 1994.

\bibitem[Potapov and Lubk(2019)]{Potapov2019}
P.~Potapov and A.~Lubk.
\newblock Optimal principal component analysis of {STEM XEDS} spectrum images.
\newblock \emph{Advanced Structural and Chemical Imaging}, 5:\penalty0 4, 2019.

\bibitem[Potapov and Engelmann(2013)]{Potapov2013}
P.~Potapov and H.-J. Engelmann.
\newblock {TEM} characterization of advanced devices in the semiconductor
  industry.
\newblock \emph{18th Conference Microscopy of Semiconducting Materials,
  Oxford}, 2013.

\bibitem[Anscombe(1948)]{Anscombe}
F.J. Anscombe.
\newblock The transformation of poisson, binomial and negativebinomial data,".
\newblock \emph{Biometrik}, 35\penalty0 (10):\penalty0 246--254, 1948.

\bibitem[Keenan and Kotula(2004)]{Keenan2004}
M.R. Keenan and P.G. Kotula.
\newblock Accounting for {P}oisson noise in the multivariate analysis of
  {TOF-SIMS} spectrum images.
\newblock \emph{Surf. Interface Anal.}, 36:\penalty0 203--212, 2004.

\bibitem[Spiegelberg et~al.(2017)Spiegelberg, Rusz, Thersleff, and
  Pelckmans]{Spiegelberg2017}
J.~Spiegelberg, J.~Rusz, T.~Thersleff, and K.~Pelckmans.
\newblock Analysis of electron energy loss spectroscopy data using geometric
  extraction methods.
\newblock \emph{Ultramicroscopy}, 174:\penalty0 14--26, 2017.

\bibitem[Kritchman and Nadler(2008)]{Kritchman2008}
S.~Kritchman and B.~Nadler.
\newblock Determining the number of components in a factor model from limited
  noisy data.
\newblock \emph{Chemometrics and Intelligent Laboratory Systems}, 94:\penalty0
  19--32, 2008.

\bibitem[Malinowski(1977)]{Malinowski1977}
E.~R Malinowski.
\newblock Theory of error in factor analysis.
\newblock \emph{Analytical Chemistry}, 49\penalty0 (4):\penalty0 606--611,
  April 1977.

\bibitem[Ghuman(2017)]{Ghuman2017}
S.S. Ghuman.
\newblock Clustering techniques- a review.
\newblock \emph{International Journal of Computer Science and Mobile
  Computing}, 5\penalty0 (5):\penalty0 524--530, 2017.

\bibitem[Fukunaga and Hostetler(1975)]{Fukunaga1974}
K.~Fukunaga and L.D. Hostetler.
\newblock The estimation of the gradient of a density function, with
  applications in pattern recognition.
\newblock \emph{IEEE Transactions on Information Theory}, 21\penalty0
  (1):\penalty0 32--40, 1975.

\bibitem[Tichonov(1963)]{Tichonov}
A.N. Tichonov.
\newblock Solution of incorrectly formulated problems and the regularization
  method.
\newblock \emph{Soviet Mathematics}, 4:\penalty0 1035--1038, 1963.

\bibitem[Craven et~al.(2017)Craven, Sawada, McFadzean, and
  MacLaren]{Craven2017}
A.J. Craven, H.~Sawada, S.~McFadzean, and I.~MacLaren.
\newblock Getting the most out of a post-column {EELS} spectrometer on a
  {TEM/STEM} by optimizing the optical coupling.
\newblock \emph{Ultramicroscopy}, 180:\penalty0 66--80, 2017.

\bibitem[Egerton(1996)]{Egerton}
R.F. Egerton.
\newblock \emph{Electron Energy-Loss Spectroscopy in the Electron Microscope}.
\newblock Plenum Press, 2nd edition, 1996.

\bibitem[Keshava and Mustard(2002)]{Keshava2002}
N.~Keshava and J.F. Mustard.
\newblock Spectral unmixing.
\newblock \emph{IEEE Signal Processing}, 19:\penalty0 44--57, 2002.

\bibitem[Halimi et~al.(2011)Halimi, Altmann, Dobigeon, and
  Tourneret]{Halimi2011}
A.~Halimi, Y.~Altmann, N.~Dobigeon, and J.-Y. Tourneret.
\newblock Nonlinear unmixing of hyperspectral images using a generalized
  bilinear model.
\newblock \emph{IEEE Trans. Geoscience and Remote Sensing}, 49\penalty0
  (11):\penalty0 4153--4162, Nov. 2011.

\bibitem[Altmann et~al.(2013)Altmann, Dobigeon, McLaughlin, and
  Tourneret]{Altmann2013}
Y.~Altmann, N.~Dobigeon, S.~McLaughlin, and J.-Y. Tourneret.
\newblock Nonlinear spectral unmixing of hyperspectral images using {G}aussian
  processes.
\newblock \emph{IEEE Trans. Signal Processing}, 61\penalty0 (10):\penalty0
  2442--2453, 2013.

\bibitem[Altmann et~al.(2014)Altmann, Dobigeon, McLaughlin, and
  Tourneret]{Altmann2014}
Y.~Altmann, N.~Dobigeon, S.~McLaughlin, and J.-Y. Tourneret.
\newblock Residual component analysis of hyperspectral images - {A}pplication
  to joint nonlinear unmixing and nonlinearity detection.
\newblock \emph{IEEE Trans. Image Processing}, 23\penalty0 (5):\penalty0
  2148--2158, 2014.

\bibitem[Halimi et~al.(2017)Halimi, {Bioucas-Dias}, Dobigeon, Buller, and
  McLaughlin]{Halimi2017}
A.~Halimi, J.~M. {Bioucas-Dias}, N.~Dobigeon, G.~S. Buller, and S.~McLaughlin.
\newblock Fast hyperspectral unmixing in presence of nonlinearity or
  mismodelling effects.
\newblock \emph{IEEE Trans. Computational Imaging}, 3\penalty0 (2):\penalty0
  146--159, 2017.

\bibitem[Cavalcanti et~al.(2019)Cavalcanti, N., and S.]{Cavalcanti2019}
Y.C. Cavalcanti, Dobigeon N., and Stute S.
\newblock Factor analysis of dynamic {PET} images: beyond {G}aussian noise.
\newblock \emph{IEEE Trans. Medical imaging}, 38:\penalty0 2231--2241, 2019.

\bibitem[T. et~al.(2019)T., Fauvel, and Dobigeon]{Uezato2019}
Uezato T., M.~Fauvel, and N.~Dobigeon.
\newblock Hyperspectral unmixing with spectral variability using adaptive
  bundles and double sparsity.
\newblock \emph{IEEE Trans. Geoscience and Remote Sensing}, 57:\penalty0
  3980--3992, 2019.

\bibitem[Lichtert and Verbeeck(2013)]{Lichert2013}
S.~Lichtert and J.~Verbeeck.
\newblock Statistical consequences of applying a {PCA} filter on {EELS}
  spectrum images.
\newblock \emph{Ultramicroscopy}, 125:\penalty0 35--42, 2013.

\bibitem[Potapov(2017)]{Potapov2017b}
P.~Potapov.
\newblock On the loss of information in {PCA} of spectrum-images.
\newblock \emph{Ultramicroscopy}, 182:\penalty0 191--194, 2017.

\bibitem[Jones et~al.(2018)Jones, Varambhia, Beanland, D., Griffiths, Ishizuka,
  Azough, Freer, K., D., M., Lozano-Perez, and Nellist]{Jones2018}
L.~Jones, A.~Varambhia, R.~Beanland, Kepaptsoglou D., I.~Griffiths,
  A.~Ishizuka, F.~Azough, R.~Freer, Ishizuka K., Cherns D., Ramasse~Q. M.,
  S.~Lozano-Perez, and P.~Nellist.
\newblock Managing dose-, damage- and data-rates in multi-frame
  spectrum-imaging.
\newblock \emph{Microscopy}, 67\penalty0 (S1):\penalty0 98--113, 2018.

\bibitem[Nadler(2008)]{Nadler2008}
B.~Nadler.
\newblock Finite sample approximation results for principal component analysis:
  a matrix perturbation approach.
\newblock \emph{Annals Statistics}, 36\penalty0 (6):\penalty0 2791--2817, 2008.

\end{thebibliography}

\end{document}


\begin{frontmatter}

\title{Extraction of physically meaningful endmembers from STEM  spectrum-images combining geometrical and statistical approaches: Supplementary material
}

\author{Pavel Potapov$^{1,*}$ and Axel Lubk$^{1,**}$}

\address{
$^1$ Leibniz Institute for Solid State and Materials Research (IFW), Dresden, Germany

$^*$ p.potapov@ifw-dresden.de

$^{**}$ a.lubk@ifw-dresden.de
    }

\end{frontmatter}

\linenumbers


\section{Weighted vs. un-weighted PCA analysis of EELS spectrum-images} 

This supplementary material demonstrates the ultimate necessity of the noise equalization   prior PCA in the case of EELS spectrum-images. Such equalization can be performed by the Anscombe transformation [37] or by the weighting procedure after Keenan and Kotula [38].

\begin{figure}[!ht]
\centering\includegraphics[width=0.9\linewidth]{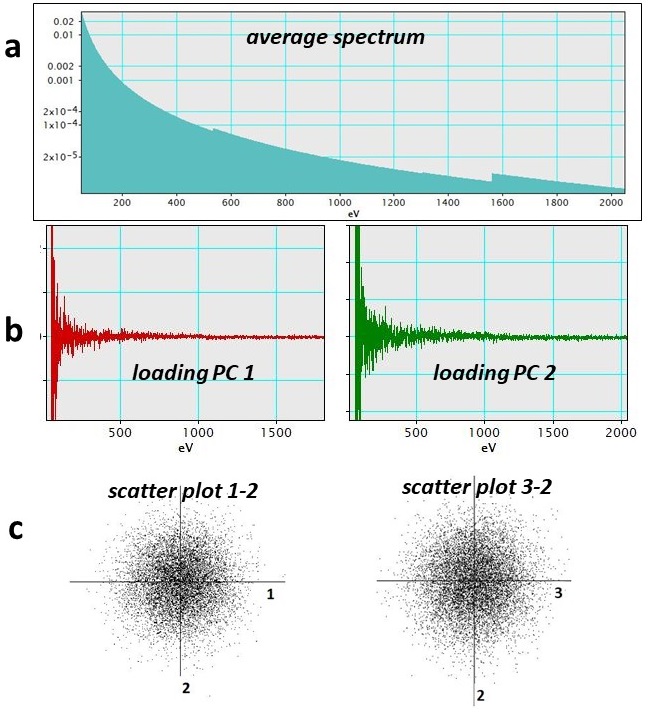}
\caption{PCA analysis of the synthetic EELS spectrum-image from [17] performed \textit{without} the weighting pre-treatment. The average spectrum (a) of the dataset  exhibits a drastic variation of the background with energy (note the logarithmic scale of the intensity). As a result, the loadings of the major principal components (b) reveal the random noise variation at the left part of spectrum instead of the variations of the EELS characteristic edges in its middle. Correspondingly, the projections from the factor space on the planes formed by  the major principal components (c) resembles the bi-variant random noise distribution.}
\label{fig:S0b}
\end{figure}

As the intensity of EEL spectra drops rapidly with increasing energy-loss, the background level at lower energies can significantly exceed the level of the useful signal at higher energies. Then PCA might completely fail as the random noise variation of the background  overrides the meaningful variations of the useful signal. Fig. S\ref{fig:S0b} demonstrates such a case for the synthetic EELS spectrum-image shown in Introduction.  It is evident that PCA is not able to extract any useful variation of the characteristic EELS edges in this case but  picks instead the random noise variations of the background at low energy-losses. This example is somewhat extreme. In most practical cases, the major variations of the EELS edges are still seen even without application of the weighting procedure. Nevertheless, the weaker principal components are typically overridden by the background variations at low energy-losses.    

The subtraction of the EELS background prior to PCA can be one possible solution to fix the problem. The background subtraction is, however, not  accurate in the presence of significant noise, in particular because the functional form of the the background is typically unknown. The better option is therefore equalizing the noise variation through the Anscombe transform or the weighting procedure as described in Section 3.1. Then the variations of the minor EELS edges are clearly visible (see Fig.1 in the main text of the article). 

\section{Anscombe transformation}

This supplementary material evaluates the PCA pre-treatment alternative to that actually applied in Section 3.1, namely Anscombe transformation [37].  

\begin{figure}[!h]
\centering\includegraphics[width=0.9\linewidth]{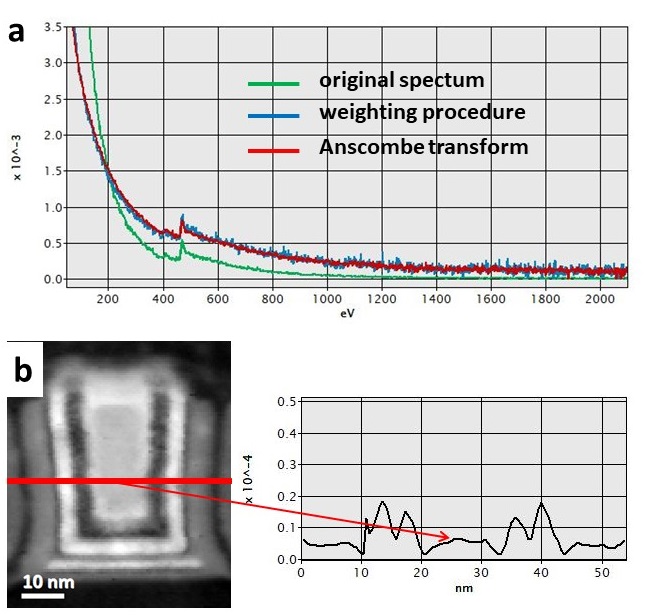}
\caption{(a) A typical EEL spectrum (green) in the spectrum-image of the CMOS device described in Section 2 and its rescaled variant in order to equalize the noise variance.  The weighing procedure (blue) and the Anscombe transformation (red) yield similar results. (b) shows the relative difference between those transformations along the representative linescan in the spectrum-image.}
\label{fig:S0}
\end{figure}

Anscombe transformation is a variance-stabilizing transformation when 
each  coordinate of a vector $\m X$ is replaced as: $x_i \rightarrow 2\sqrt{x_i+\nicefrac{3}{8}}$. With respect to an EELS spectrum-image, this means that the counts $I$ at each energy channel of a spectrum are replaced by $2 \sqrt{I +\nicefrac{3}{8}}$. Fig. S\ref{fig:S0} demonstrates that the results of the weighting procedure described in Section 3.1 and the Anscombe transformation are very similar\footnote{For the sake of comparison, the intensity scale of the curves was tuned  such as the areas under the curves were equal.}. The relative difference between two pre-treatments is typically about $10^{-5}$ as seen from Fig. S\ref{fig:S0}b.

\section{Endmembers extraction using Joint Bayesian Linear Unmixing (BLU) algorithm [25,27]} 

This supplementary material explores the endmember analysis alternative to that in Section 3.4.1, namely joint Bayesian endmember extraction after Dobigeon et al. [25,27].

The dataset described in Section 2 of the main text was processed by the MatLab-based  \footnote{http://dobigeon.perso.enseeiht.fr/applications/app\_EELS.html} Bayesian Linear Unmixing (BLU) package  [25,27]. For the purpose of comparison with the algorithm of the present paper, the target number of endmembers was chosen as 7.  This implies a dimensionality reduction of the factor space in accordance with  the chosen number of endmembers, although a PCA was not explicitly applied to the dataset. The positions of the endmembers were first estimated by VCA and then refined by the Bayesian inference. The number of Markov chains iterations in the refinement was 100. To check the reproducibility of the treatment, the procedure was repeated several times.  Although the initial VCA positions varied quite noticeably, the refined positions were almost identical, which proved good convergence properties of the algorithm. 

\begin{figure}[!ht]
\centering\includegraphics[width=1.0\linewidth]{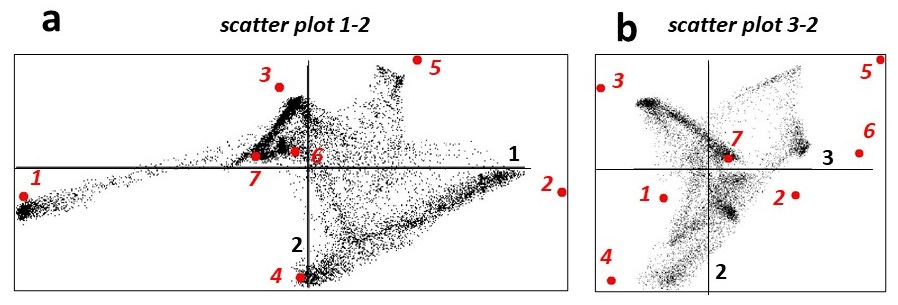}
\caption{The 7 endmember positions found  by applying the BLU algorithm after Dobigeon et al. [25,27] shown in projections on planes formed by (a) the 1st and 2nd and  (b) 2nd and 3rd principal components. The endmembers are labeled in the same sequence as the endmembers shown in Fig. 6.}
\label{fig:S1}
\end{figure}

Fig. S\ref{fig:S1} shows the positions of the obtained endmembers in factor space. Although their configuration generally reproduces that in Fig. 6, the locations of most   endmembers are noticeably shifted. We should stress that the algorithm attempts to estimate the noise level from the data distribution, which is challenging given the large and inhomogeneous spread of data points withing the variation polyhedron in this data set.  In contrast, the algorithm described in Section 3 of the manuscript assumes a fixed noise level estimated from the residual matrix according to Malinowski [40], which could be the reason for the observed difference. 

\begin{figure}[!ht]
\centering\includegraphics[width=1.0\linewidth]{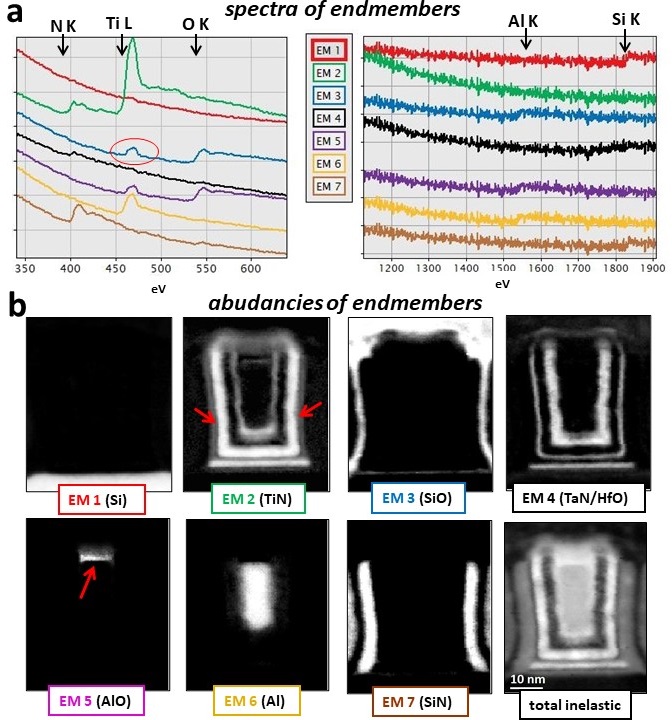}
\caption{Spectra (a) and  abundances maps (b) of the endmembers obtained after Dobigeon et al. [25,27] The spectra are shown in the lower (left) and higher (right) energy regions. In comparison to Fig. 7, an extra Ti L peak is observed in the spectrum of endmember 3 (outlined by red). Also, the abundances maps of endmember 2 and 5 show discrepancies with the corresponding maps in Fig. 7 denoted by the red arrows. The maps are displayed with the contrast parameters equal to those in Fig. 7.}
\label{fig:S2}
\end{figure}

Nevertheless, the obtained endmember spectra and abundances fit reasonably well to those reported in Section 3 as  seen from Fig. S\ref{fig:S2}. Still, the following discrepancies with the results in Fig. 7 are apparent: i) a small Ti L peak in the spectrum of  endmember 3 (supposed to be SiO), ii) a weak extra layer in the map of endmember 2 (supposed to be TiN), iii) a shrunken area occupied by endmember 5 (supposed to be AlO). These features represent most probably artifacts of the treatment.

On the other hand, the abundances maps in Fig. \ref{fig:S2} are noticeably less noisy than those in Fig. 7. We attribute this to the stronger reduction of the factor space dimensionality  executed in the course of the treatment.

\section{Endmembers determination after Spiegelberg et al. [39]} 

This supplementary material explores the endmember analysis alternative to that in Section 3.4.1, namely scattered VCA after Spiegelberg et al. [39].

\begin{figure}[!ht]
\centering\includegraphics[width=1.0\linewidth]{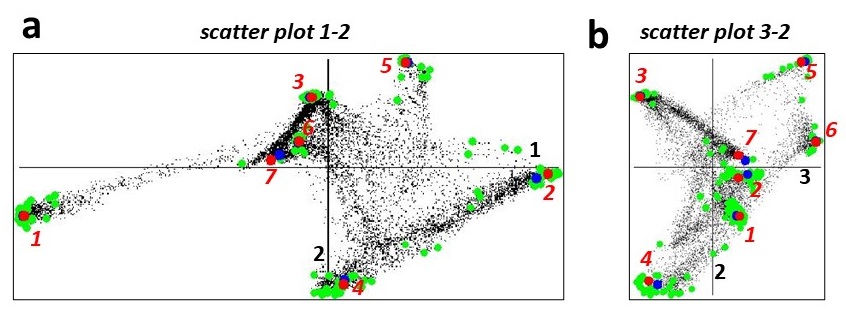}
\caption{The VCA endmember positions (green spots) generated according to Spiegelberg et al. [38] shown in the 2D projections spanned by (a) the 1st and 2nd and  (b) 2nd and 3rd principal components. The final 7 endmembers obtained by clustering are shown as red spots and numbered as in Fig.6a,b. For the comparison purpose, the endmember positions derived in Section 3.4.2 are shown as blue spots.}
\label{fig:S3}
\end{figure}

For the PCA-reduced dataset described in Section 3.1, we deliberately generated 560 endmembers  according to the algorithm [39]. This number corresponded to the number of endmembers generated by our algorithm in Section 3.4.1. As the algorithm [39] is expected to be sensitive to outliers, we carefully striped the dataset from extreme variations like X-ray spikes prior to the treatment. Nevertheless, Table \ref{tab:S1} demonstrates that a large number of obtained endmembers are redundant and only 12 percent of them are unique. Despite of that, Fig. S\ref{fig:S3} demonstrates that the positions of the endmembers are distributed similar to those obtained in Section 3.4.1 (see Fig. 6 a,b) although the borders of a multi-dimensional polyhedron are defined  less clearly due to the reduced number of  unique points. Nevertheless, if we fix the final number of endmembers to 7 (same as in Section 3.4.1), the clustering algorithm performs quite well yielding the  endmember positions very close to those in Fig. 6. 

\begin{table}[h]
\centering
\begin{tabular}{| l l | l l | l l |}
\hline
\textbf{number of} & \textbf{times} & \textbf{number of} & \textbf{times}\\
\textbf{endmembers} & \textbf{repeated} & \textbf{endmembers} & \textbf{repeated}\\
\hline
1 & 44 & 4 & 7 \\
1 & 38 & 7 & 6 \\
1 & 14 & 8 & 5 \\
2 & 12 & 12 & 4 \\
1 & 11 & 15 & 3 \\
4 & 10 & 39 & 2 \\
2 & 9 & 66 & 1 \\
3 & 8  &  & \\
\hline
\end{tabular}
\caption{Coincidence among endmembers generated according to the algorithm [39]. The left column lists the number of data points, which were identified as endmembers several times as indicated in the right column. Only 66 from 560 generated points represented unique endmembers.}
\label{tab:S1}
\end{table}

\begin{figure}[!ht]
\centering\includegraphics[width=1.0\linewidth]{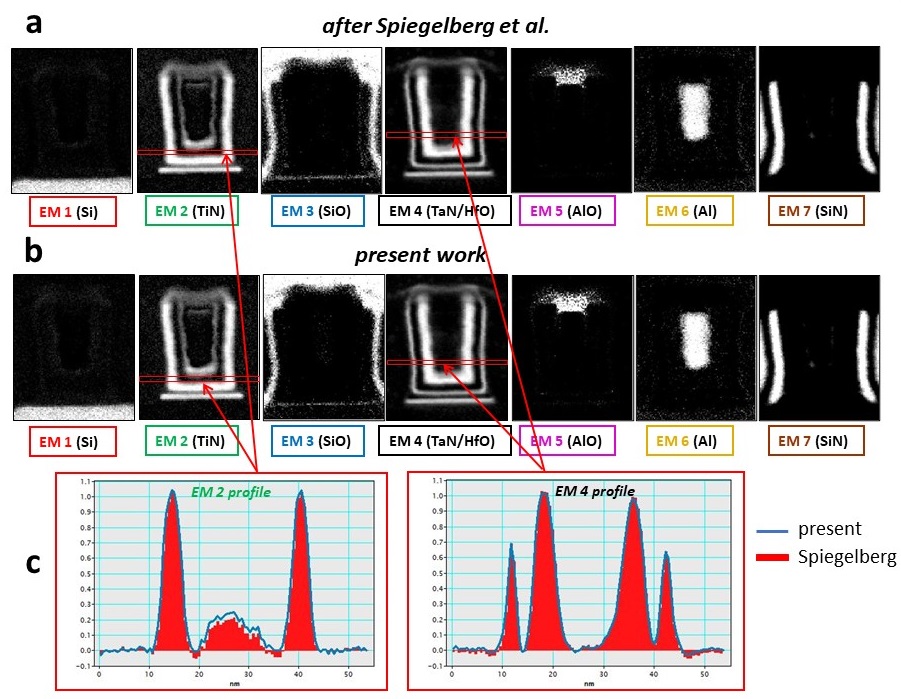}
\caption{Comparison of the abundance maps obtained by the  algorithm of (a) Spiegelberg et al. [38] and (b) by the method in the present work. The contrast is set identically in both series. The tiny differences can be seen only in the line profiles shown in (c). }
\label{fig:S4}
\end{figure}

The spectra of the final endmembers are similar to those in Fig.7, thus they are not shown here. Also the abundance maps (Fig. S\ref{fig:S4}a,b) show close proximity with the results in Section 4. The tiny quantitative differences can only be captured when comparing profiles across the maps (Fig. S\ref{fig:S4}c).  Accordingly, the abundances are eventually shifted to slightly negative values in areas where they are expected to be zero. These small inaccuracies seems to appear because the algorithm [38] does not account for the noise as the BLU algorithm [25] and the algorithm suggested in the present work.

\section{STEM EDX spectrum-image of a CMOS transistor} 

A CMOS transistor identical to that described in Section 2 was   scanned for 6 min in the same (S)TEM microscope while  registering the EDX signal by a 4-window SDD detector. The STEM EDX spectrum-image consisting of $244 \times 336$ probe positions and 1200 energy channels was subjected to the PCA treatment as described in [35], which reduced the dimensionality of the factor space down to 7. Then, the algorithm described in Section 3   was applied to extract physically relevant endmembers.

\begin{figure}[!ht]
\centering\includegraphics[width=1\linewidth]{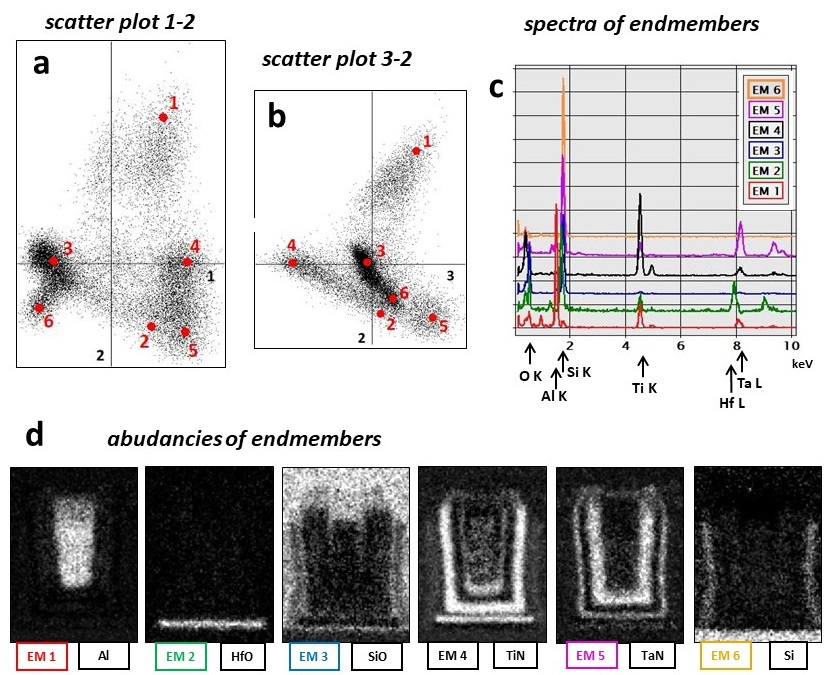}
\caption{Endmember analysis for the STEM EDX spectrum-image of the CMOS transistor. (a) and (b) are the scatter plots of the 1st \textit{vs} 2nd and 2nd \textit{vs} 3rd principal components respectively with the positions of the retrieved endmembers indicated; (c) shows the spectra of the endmembers and (d) are their abundance maps.}
\label{fig:S5}
\end{figure}

Only 6 major endmembers were found to exhibit a clear physical meaning. Their positions in the factor space as well as the spectra and the abundance maps are shown in Fig. S\ref{fig:S5}. The EDX results are coherent with the EELS results in Fig. 7 of the main text when taking into account the different sensitivity of the two methods to various ionisation shells. Namely, the Hf signal is well visible and separable from the Ta one in EDX but not in EELS. In contrast, the EDX sensitivity to light element is much weaker than that for EELS, therefore the AlO compound was not identified as a distinct endmember in the EDX spectrum-image. We stress that the scanning was executed at a relatively low beam current. An increase of the beam current  or acquisition time allows for the reliable  separation of the Al and the AlO compound.  The SiN compound is missing in Fig. S\ref{fig:S5} because the scanning was executed over a slightly  narrower area than that in Fig. 7.

